\documentclass[sigconf]{acmart}

\AtBeginDocument{%
  }

\setcopyright{acmcopyright}
\copyrightyear{2024}
\acmYear{2024}
\setcopyright{licensedothergov}\acmConference[CCS '24]{Proceedings of the
2024 ACM SIGSAC Conference on Computer and Communications
Security}{October 14--18, 2024}{Salt Lake City, UT, USA}
\acmBooktitle{Proceedings of the 2024 ACM SIGSAC Conference on Computer
and Communications Security (CCS '24), October 14--18, 2024, Salt Lake City,
UT, USA}
\acmDOI{10.1145/3658644.3690369}
\acmISBN{979-8-4007-0636-3/24/10}

\usepackage{xspace}
\usepackage{nicefrac}
\usepackage{subcaption}
\usepackage{multirow}
\usepackage{amsfonts}
\usepackage{amsmath}
\usepackage{amsthm}
\usepackage{numprint} % problem
\usepackage{hyperref}
\usepackage{enumitem}
\setlist[itemize]{leftmargin=*}
\usepackage{makecell}
\usepackage{color}
\hypersetup{
    colorlinks=false,
    pdfborder={0 0 0},
}
\DeclareMathAlphabet{\mathcal}{OMS}{cmsy}{m}{n}

\newcommand{\ournameNoSpace}{\emph{Phantom}}
\newcommand{\paperTitle}{Phantom: Untargeted Poisoning Attacks on Semi-Supervised Learning (Full Version)*}
\newcommand{\paperTitleNoLineBreak}{\paperTitle}

\newcommand{\adversaryNoSpace}{\ensuremath{\mathcal{A}}}
\newcommand{\adversary}{\adversaryNoSpace\xspace}
\newcommand{\adversaryGen}{\adversaryNoSpace's\xspace}
\newcommand{\labeledDataset}{\ensuremath{\mathcal{X}}\xspace}
\newcommand{\datasetLabeled}{\labeledDataset}
\newcommand{\datasetUnlabeled}{\ensuremath{\mathcal{U}}\xspace}
\newcommand{\unlabeledDataset}{\datasetUnlabeled}

\newcommand{\ourname}{\ournameNoSpace\xspace}
\newcommand{\ournameAttackNew}{\ourname attack\xspace}
\newcommand{\ournameGen}{\ournameNoSpace's\xspace}

\newcommand{\sota}{state-of-the-art\xspace}
\newcommand{\etal}{\emph{et~al.}\xspace}

\newcommand{\cifarNoSpace}{\mbox{CIFAR-10}}
\newcommand{\cifar}{\cifarNoSpace\xspace}
\newcommand{\svhn}{SVHN\xspace}
\newcommand{\stl}{\mbox{STL-10}\xspace}
\newcommand{\mnist}{MNIST\xspace}
\newcommand{\nDatasets}{six\xspace}
\newcommand{\gtsrb}{GTSRB\xspace}
\newcommand{\imagenet}{ImageNet\xspace}

\newcommand{\platformNoSpace}{\ensuremath{\mathcal{P}}}
\newcommand{\platform}{\platformNoSpace\xspace}
\newcommand{\victimNoSpace}{\ensuremath{\mathcal{V}}}
\newcommand{\victim}{\victimNoSpace\xspace}
\newcommand{\victimGen}{\victimNoSpace's\xspace}

\newcommand{\challengeNew}[1]{C#1}

\newcommand{\fixmatchNoSpace}{FixMatch}
\newcommand{\mixmatchNoSpace}{MixMatch}
\newcommand{\udaNoSpace}{UDA}
\newcommand{\fixmatch}{\fixmatchNoSpace\xspace}
\newcommand{\mixmatch}{\mixmatchNoSpace\xspace}
\newcommand{\uda}{\udaNoSpace\xspace}
\newcommand{\sect}{Sect.~}
\newcommand{\appSect}{App.~}

\newcommand{\scaleTable}[1]{\scalebox{.9}{#1}}

\newcommand\extrafootertext[1]{%
    \bgroup
    \renewcommand\thefootnote{\fnsymbol{footnote}}%
    \renewcommand\thempfootnote{\fnsymbol{mpfootnote}}%
    \footnotetext[0]{*~#1}%
    \egroup
}

\begin{document}

%\fancyhead[LO]{\sffamily\footnotesize G}
\title[\paperTitleNoLineBreak]{\paperTitle}
%\def\shorttitle {\paperTitleNoLineBreak}


\author{Jonathan Knauer}
\email{jonathan.knauer@stud.tu-darmstadt.de}
\affiliation{
  \institution{Technical University of Darmstadt}
  \country{Germany}
}
\author{Phillip Rieger}
\email{phillip.rieger@trust.tu-darmstadt.de}
\affiliation{%
  \institution{Technical University of Darmstadt}
  \country{Germany}
}

\author{Hossein Fereidooni}
\email{hossein.fereidooni@kobil.com}
\affiliation{%
  \institution{KOBIL GmbH}
  \country{Germany}
}

\author{Ahmad-Reza Sadeghi}
\email{ahmad.sadeghi@trust.tu-darmstadt.de}
\affiliation{%
  \institution{Technical University of Darmstadt}
  \country{Germany}
}


\begin{abstract}
    Deep Neural Networks (DNNs) can handle increasingly complex tasks, albeit they require rapidly expanding training datasets. Collecting data from platforms with user-generated content, such as social networks, has significantly eased the acquisition of large datasets for training DNNs. Despite these advancements, the manual labeling process remains a substantial challenge in terms of both time and cost. In response, Semi-Supervised Learning (SSL) approaches have emerged, where only a small fraction of the dataset needs to be labeled, leaving the majority unlabeled. However, leveraging data from untrusted sources like social networks also creates new security risks, as potential attackers can easily inject manipulated samples. Previous research on the security of SSL primarily focused on injecting backdoors into trained models, while less attention was given to \mbox{the more challenging untargeted poisoning attacks.}\\
\noindent In this paper, we introduce \ourname, the first untargeted poisoning attack in SSL that disrupts the training process by injecting a small number of manipulated images into the unlabeled dataset. Unlike existing attacks, our approach only requires adding few manipulated samples, such as posting images on social networks, without the need to control the victim. \ourname causes SSL algorithms to overlook the actual images' pixels and to rely only on maliciously crafted patterns that \ourname superimposed on the real images.
We show \ournameGen effectiveness for \nDatasets different datasets and 3 real-world social-media platforms (Facebook, Instagram, Pinterest). Already small fractions of manipulated samples (e.g., 5\%) reduce the accuracy of the resulting model by 10\%, with higher percentages leading to a performance comparable to a naive classifier. Our findings demonstrate the threat of poisoning user-generated content platforms, rendering them unsuitable for SSL in specific tasks.

\end{abstract}


%\ccsdesc[500]{Computer systems organization~Embedded systems}

%\keywords{datasets, neural networks, gaze detection, text tagging}

\maketitle
%\mbox{ }\\\vspace{-18.5cm}\mbox{ }\\{\color{red}Target Conference ACM CCS 2023\hfill Revision 1\\ Deadline: 19 January, 2023. \hfill\\ Page Limit: 12 pages main part}\vspace{16cm}\\% 
\section{Introduction}
\label{sect:intro}
\extrafootertext{© Knauer 2024. This is the author's version of the work. It is posted here for
your personal use. Not for redistribution. The definitive version was published
in {CCS2024}, https://doi.org/10.1145/3658644.3690369."}

\noindent Deep Neural Networks (DNNs) have been rapidly evolving with spectacular abilities, such as \mbox{DALL-E}~\cite{ramesh2022hierarchical}, DeepFakes~\cite{nguyen2022deep}, and the chatbot ChatGPT~\cite{chatgpt}, to name a few. Simultaneously, DNNs are increasingly used also for safety-critical domains, such as self-driving cars, necessitating exceptionally high reliability and precision. However, with the growing complexity of these tasks, also the size of the DNNs and the number of trainable parameters grow significantly. To effectively train these DNN models, the required amount of training data is growing rapidly. One possible source of large amounts of training data is the Internet, where large amounts of user-generated content are available on platforms such as Instagram (image data), Reddit (text data), and YouTube (video data). Thus, the data that were uploaded on these platforms became a high-value asset, as different actors, such as companies, can use these data to train DNNs.\\
However, labeling the data remains a significant practical challenge in terms of cost and time. To address this issue, recent research has focused on developing algorithms that either avoid the need for labeling entirely or minimize the number of samples that require manual annotation. \\
Notably, recent news about the achievements of DNNs focus on Large Language Models (LLMs) that are trained in early training phases on text sequences in a self-supervised manner. However, these models still require labeled training data for later supervised training phases~\cite{openAI23gpt,openAI24FineTuning,Perrigo2023}. Furthermore, other applications, such as image recognition or those developed by stakeholders such as small companies or academic institutions~\cite{fereidooni2023authentisense}, typically do not rely exclusively on self-supervised training methods. This shows the ongoing need for labeled training data, despite significant progress in self-supervised learning approaches.

\noindent \textbf{Semi-Supervised Learning.} 

Semi-Supervised Learning (SSL) algorithms avoid the problem of time-intensive labeling the obtained large dataset by minimizing the number of labeled samples that are required for the training algorithm. SSL algorithms utilize two datasets, a small labeled dataset and a large unlabeled dataset. The model is initially trained only on the labeled dataset, while the current model is simultaneously used to predict the labels for samples from the unlabeled dataset. Once the model predicts the labels for some samples with high confidence, these samples are used with these guessed labels for further training.
\noindent Current \sota algorithms for SSL include MixMatch~\cite{berthelot2019mixmatch}, UDA~\cite{xie2020unsupervised}, and FixMatch~\cite{sohn2020fixmatch}. These algorithms are highly effective at leveraging only a small number of labeled samples to achieve a performance comparable to fully supervised methods. For example, on the \cifar benchmark dataset, labeling a only 40 out of \numprint{50000} samples is sufficient to achieve decent accuracy levels. SSL algorithms allow leveraging large unlabeled datasets, e.g., downloaded from social media platforms, while avoiding the costly and time-consuming process of manual labeling. 

\noindent\textbf{Attacks on SSL.} 
However, despite the benefits of leveraging large datasets from platforms with user-generated content, using untrusted data also creates new attack vectors. If a party like a company utilizes the user-generated data of such a platform, an attacker can easily upload maliciously crafted images to the platform to manipulate DNNs that are trained on this data. 
Recent research proposed attacks that insert manipulated samples into the unlabeled training data to inject a backdoor into the trained model. The backdoor causes the trained DNN to misbehave in a well-defined manner, such as classifying selected images as belonging to a different class chosen by the attacker~\cite{carlini2021poisoning,connor2022rethinking}. Previous work developed untargeted poisoning attacks that reduce the trained model's accuracy by manipulating the labeled dataset~\cite{franci2022influence,liu2019unified}. However, to manipulate the labeled samples, a strong control over the attacked victim is required. Thus, this would also enable the adversary to interfere directly with the training process. 

\noindent While existing untargeted poisoning attacks assume access to the labeled dataset, to the best of our knowledge, no untargeted attack has been proposed that manipulates the unlabeled data to disrupt the training process. Compared to backdoor attacks that manipulate the unlabeled dataset~\cite{carlini2021poisoning,connor2022rethinking}, untargeted poisoning attacks pose significant challenges. In backdoor attacks, the DNN model can achieve both objectives simultaneously, a decent performance on the benign data and injecting the backdoor. Backdoor attacks only add an additional, undesired function to the DNN. Consequently, the poisoned data only need to provide examples for this additional function such that if the DNN is trained on these data, it will also learn this function. In contrast, untargeted poisoning attacks, which aim to compromise the performance on benign test data, present a more challenging scenario as here both above-mentioned objectives will be in direct conflict with each other. Thus, one of the challenging aspects when designing untargeted attacks is the necessity for a relatively small number of samples to overrule the impact of a large number of benign samples. 

\noindent Another challenge is the lack of information about the SSL algorithm, DNN architecture, or used model weights. As the manipulated samples must be crafted and uploaded to the platform with user-generated content before the actual training starts, the attack has to be executed blindly. Moreover, generating inaccurate labels for the manipulated data remains another challenge since the labels for the unlabeled dataset are obtained by the SSL algorithm during the training.

\begin{figure}[tb]
    \centering
    \includegraphics[width=.85\linewidth,trim={0 0.77cm 0cm 1.1cm},clip]{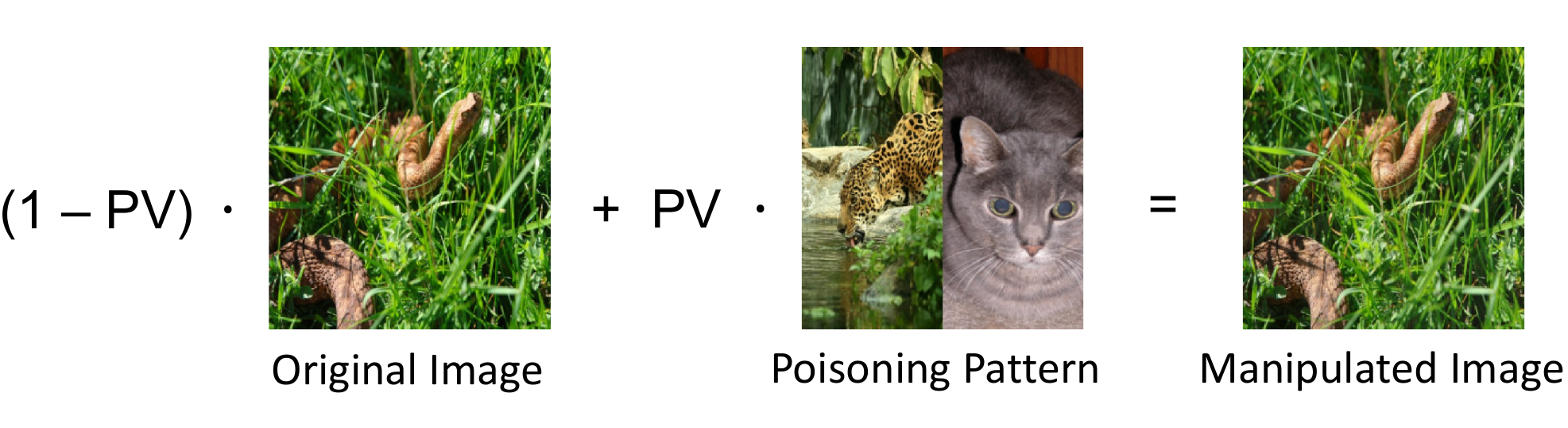}
    \caption{Example for poisoned image construction as a combination of the unpoisoned image and the Poisoning Pattern, weighted by the Pattern Visibility (PV) parameter, here 0.1.}
    \label{fig:poisonedConstruction}
\end{figure}

\noindent\textbf{Our goal and contributions.}
In this paper, we propose \ourname, the first blind untargeted poisoning attack on SSL that disrupts the training by manipulating the unlabeled dataset. By adding a small number of manipulated samples to the unlabeled dataset, such as uploading manipulated images to a social network, \ourname significantly reduces the performance of the trained model. To effectively disturb the SSL algorithm, we incorporate a hidden pattern (a \ourname) into the manipulated samples, causing the algorithms to overlook the actual content (e.g., original pixels) and rely only on hidden patterns that we superimposed on the original content for guessing the unlabeled samples' labels. This causes the SSL algorithm to mislabel the manipulated samples and, in turn, negatively impacts the training and prevents the utilization of benign samples. An example, using images as a demonstration, is illustrated in Fig.~\ref{fig:poisonedConstruction}.
Here, the DNN uses a part of the poisoning pattern, such as the tiger or cat, to predict the label while primarily using the snake for loss calculation and parameter adaption.

\noindent It is worth noting that even with a small percentage of manipulated data, such as 5\%, the accuracy of the resulting model can be reduced by 10\%. We demonstrate that untargeted poisoning attacks can be made without making any assumptions, such as strong control or the ability to monitor the victim. Thereby, we show that untargeted poisoning attacks pose a practical and realistic threat.

\noindent In summary, our contributions include:
\begin{itemize}

    \item We propose \ourname, the first untargeted poisoning attack that disrupts the training of SSL 
    algorithms and reduces the model's utility by only adding few samples to the unlabeled dataset without requiring any knowledge of the SSL algorithm, DNN structure, or hyperparameters. In contrast to backdoor attacks, where the benign and backdoor objectives can be achieved in parallel, \ourname overrules the benign majority of samples to reduce the model's utility. \ourname poisons sources for unlabeled data, making them unusable for SSL training in certain tasks~(\sect\ref{sect:approach-design}).%
    
    \item We design a scheme for manipulating samples that takes advantage of the model's overfitting on the labeled dataset in the initial stages of training. This causes the SSL algorithms to ignore the original sample and entirely rely on superimposed patterns, thus mislabeling the sample~(\sect\ref{sect:approach}). 
    \item We conduct an extensive evaluation and demonstrate that already 5\% of manipulated data can lead to a significant reduction in the performance of the trained model by 10\%. We effectively show the efficacy of \ourname across a range of \sota SSL algorithms, datasets, and parameter configurations and its resistance to data augmentation techniques and countermeasures against adversarial examples. Additionally, we thoroughly examine the attack's impact on the model's behavior in-depth using explainable AI techniques~(\sect\ref{sect:eval}).
    
    \item To demonstrate the risks posed by the \ournameAttackNew and the need for robust SSL algorithms, we show its effectiveness through a real-world case study. We utilize data extracted from three major social networks~—~Facebook, Instagram, and Pinterest — which host user-generated content and are plausible sources for SSL training data (\sect\ref{sect:eval-casestudy}).
   
\end{itemize}

\section{Background}

\noindent In the following sections, we provide an overview of the necessary background for the rest of the paper. In \sect\ref{sect:background-ssl}, we introduce several \sota SSL algorithms (\mixmatch~\cite{berthelot2019mixmatch}, \uda~\cite{xie2020unsupervised}, and \fixmatch~\cite{sohn2020fixmatch}), which will be used in this paper. These algorithms are chosen for their popularity and effectiveness in various SSL tasks. Later, we describe and categorize different attacks that target DNN models during the training phase (\sect\ref{sect:background-poisoning}). Further, in \appSect\ref{sect:background-augmentation}, we describe an overview of data augmentation techniques that are commonly employed in the field of SSL.

\subsection{Semi-Supervised Learning}
\label{sect:background-ssl}

\noindent Semi-Supervised Learning (SSL) is a class of algorithms that utilize a combination of a small labeled dataset, denoted as $\datasetLabeled$, and a much larger unlabeled dataset, denoted as $\unlabeledDataset$, where $|\datasetLabeled| \ll|\unlabeledDataset|$, to train a Machine Learning model for classification tasks, such as a Deep Neural Network (DNN). In contrast, non-SSL algorithms require the entire dataset to be labeled, which can be costly in terms of time and expertise. SSL algorithms only require a small portion of the dataset to be labeled. The labeled part \datasetLabeled is used for the initial training of the DNN and to make educated guesses about the labels of the unlabeled data during training. In this work, we exemplary focus on the algorithms \mixmatch~\cite{berthelot2019mixmatch}, \uda~\cite{xie2020unsupervised}, and \fixmatch~\cite{sohn2020fixmatch}, as these have been shown to achieve the best performance for image applications~\cite{carlini2021poisoning}.\\
\noindent\textbf{\mixmatchNoSpace} generates for each unlabeled sample $u\in\unlabeledDataset$, i.e., an image, multiple augmented versions and averages the predictions of the current model to obtain a guess for the probability distribution for the label of $u$. For a sample $u\in\unlabeledDataset$ the probability distribution $p_u$ is represented as a vector. For each possible label (or category) $c\in \mathcal{C}$, the respective element $p_{u,c}$ indicates the probability that $u$ belongs to $c$. This guessed probability distribution is then sharpened based on a temperature parameter $T$ as:

\begin{equation}
    \text{Sharpen}(p_u,T)_i=\frac{p_{u,i}^{\nicefrac{1}{T}}}{\sum\limits_{c\in \mathcal{C}}{p_{u,c}^{\nicefrac{1}{T}}}}
\end{equation}
The loss $\mathcal{L}=\mathcal{L_X} + \lambda_{\unlabeledDataset}\mathcal{L_U}$ is then defined as the sum of the binary-cross-entropy loss $\mathcal{L_X}$ for the labeled training data $\labeledDataset$, and the distance $\mathcal{L_U}$ between the predictions for the augmented unlabeled samples $u\in\unlabeledDataset$ to the sharpened guessed label. The unlabeled data's loss is in addition weighted by a hyperparameter $\lambda_{\unlabeledDataset}$~\cite{berthelot2019mixmatch}.\\
\noindent\textbf{\uda} 

also generates multiple augmented versions of each unlabeled sample to guess the correct label. However, \uda uses different types of augmentations, one for guessing the label and another for calculating the loss. To guess the label, \uda applies weak augmentation to the sample, while for the loss calculation, it uses strong augmentation, causing stronger deformations of the images. The loss for the unlabeled example is based on the prediction of the strongly augmented version. The algorithm optimizes the parameters of the DNN to make the prediction of the strong augmented sample match the label that was predicted for the weakly augmented versions. Additionally, \uda only uses samples where the confidence for the prediction is higher than a configurable threshold~\cite{xie2020unsupervised}.

\noindent\textbf{\fixmatch} 

builds on \uda by also using different augmentation types for guessing the label and for minimizing the loss. However, unlike \uda, \fixmatch does not sharpen the probability distribution of the guessed label. Instead, it selects the label with the highest score and uses a one-hot encoding of this label~\cite{sohn2020fixmatch}.

\subsection{Poisoning Attacks}
\label{sect:background-poisoning}
\noindent Attacks that alter the training data of a DNN are often referred to as poisoning attacks~\cite{carlini2021poisoning,fang2020local,feng2022unlabeled,franci2022influence,zhang2020online}. Such attacks can be classified into the following categories:\\
\textbf{Targeted Poisoning:} 

In this type of attack, the attacker aims to manipulate the DNN's predictions towards a specific target class for certain samples. Thus, targeted attacks inject a hidden function within the DNN. Some literature~\cite{carlini2021poisoning} subdivides this attack type into two subcategories. The first subcategory involves attacks that cause the DNN to misclassify only a fixed set of samples. Attacks in the second subcategory cause the DNN to misclassify all samples with a specific trigger, such as the presence of a certain colored patch or object in an image. \\
\textbf{Untargeted Poisoning:} 

This category of attacks aims to degrade the overall performance of the DNN on all input samples by reducing its accuracy to that of a simple or naive model. This is different from targeted attacks, where the goal is to make the DNN misclassify only specific samples. The challenge in untargeted attacks is that their objective is to negatively impact the performance of the DNN on all data, including non-poisoned samples. In targeted attacks, the DNN can have a high accuracy on non-poisoned data and still misclassify the targeted samples. Thus, in targeted attacks both, the objective from benign data, to learn recognizing images, as well as the objective of the targeted attacks, of misclassifying the backdoor samples, can be achieved. In comparison, in untargeted attacks, either the attack is successful or the model achieves a decent performance on the benign training data. Therefore, the manipulated samples here need to counteract the benign data.

\section{Problem Setting}
\label{sect:problem}
\noindent In the following, we describe the considered system (\sect\ref{sect:problem-system}), define the threat model (\sect\ref{sect:problem-adversary}), and describe the challenges that untargeted attacks face in SSL (\sect\ref{sect:problem-requ}).

\subsection{System Setting}
\label{sect:problem-system}

\begin{figure}[tb]
    \centering
    
    \begin{subfigure}[b]{0.4\textwidth}
        \centering
        \includegraphics[width=0.85\textwidth,trim={5.725cm 7.65cm 8.4cm 5.95cm},clip]{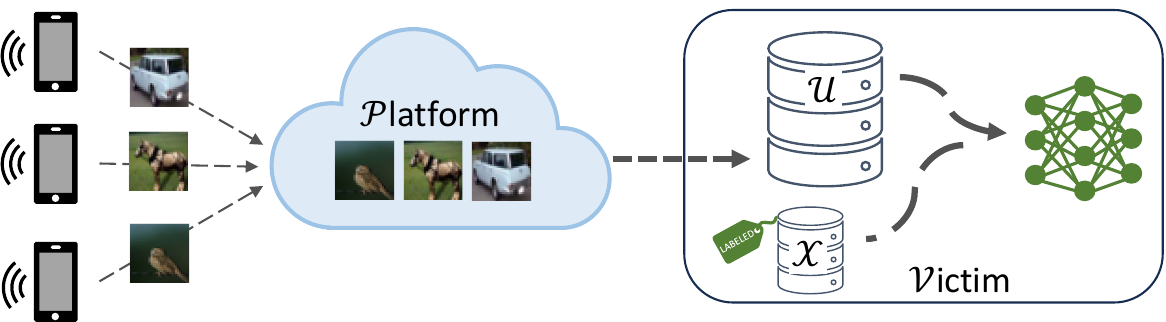}
        \caption{Without Attack}
  	    \label{fig:overview:benign}
    \end{subfigure}
    \hfill
    \begin{subfigure}[b]{0.4\textwidth}
         \centering
         \includegraphics[width=0.85\textwidth,trim={5.725cm 7.65cm 8.4cm 3.4cm},clip]{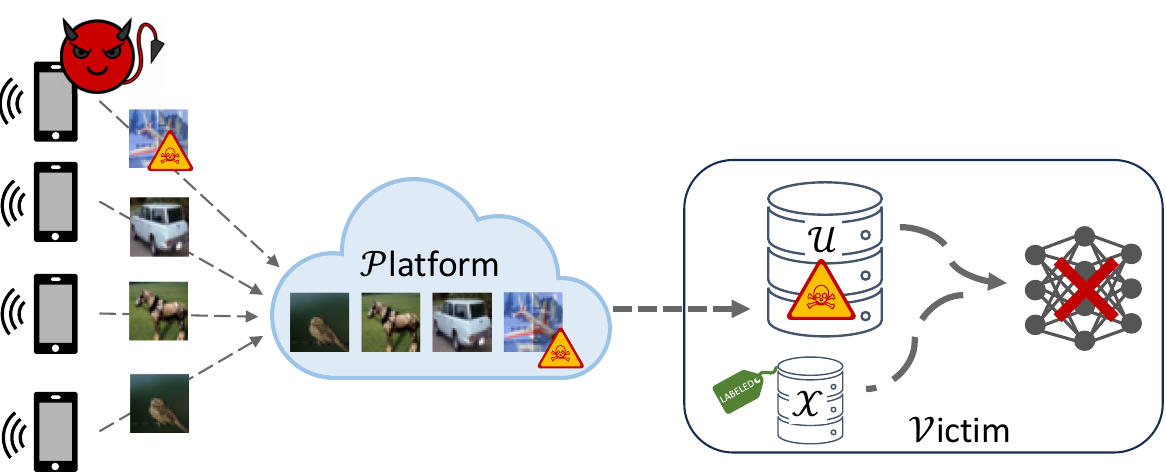}
         \caption{\ourname}
  	     \label{fig:overview:attack}
    \end{subfigure}
     
    \caption{Overview of the considered system where the \victimNoSpace ictim downloads data from a \platformNoSpace latform with user-generated content to obtain an unlabeled dataset (\unlabeledDataset) that is used with a small labeled dataset (\datasetLabeled) to train a DNN} 
    \label{fig:overview}
\end{figure}

\noindent We consider a system where anonymous clients upload user-generated content, such as images, to an internet platform, denoted as \platform. Real-world examples of such platforms are social media websites, such as Instagram, Facebook, or Pinterest. The platform \platform may perform modifications on the content, such as compressing images to reduce their size. Afterward, the victim \victim accesses the publicly available content on \platform to obtain a large dataset, referred to as \unlabeledDataset, for training a DNN using a SSL algorithm of their choice. In addition to \unlabeledDataset, \victim also uses a small dataset, referred to as \datasetLabeled, that has been labeled manually. Since the process of labeling is time-consuming and requires human effort, it is assumed that $|\datasetLabeled|\ll|\unlabeledDataset|$. Figure~\ref{fig:overview} illustrates the considered system.

\subsection{Threat Model}
\label{sect:problem-adversary}
\noindent We consider a weak adversary \adversary, whose only capability is the ability to upload data to a public platform \platform hosting user-generated content. \adversary aims to prevent the victim \victim from training a model on data that is uploaded to \platform. The purposes of such attacks might include disrupting \victim, for instance, if \adversary and \victim are competing entities, or preventing parties from utilizing the data that is available on \platform. In the latter case, \adversary could have commercial goals, e.g., revealing the poisoned samples only after \victim pays a ransom. \adversaryGen objective can be formulated as follows: \\

\noindent\textbf{Adversary's Objective} is to significantly\footnote{There exists no precise threshold which performance drop can be considered to be significant as this depends on the domain. In safety-critical applications, such as self driving cars, even a drop by 1\% already can make the model unusable, while in applications recognizing pictures on smartphones 1\% can be tolerated. We elaborate on this in \sect\ref{sect:discussion-impact}.} decrease the utility of DNN models that are trained in a semi-supervised manner on \unlabeledDataset.\\

\noindent As \platform is a public platform with user-generated content, it is sufficient for \adversary to control at least one client, such as a mobile phone or a computer. We, therefore, assume that \adversary can use such a client to upload content to \platform. Also, we assume that \adversary\xspace can make educated guesses for some labeled samples from \datasetLabeled. It is important to note that \adversary does not need to guess the samples in \datasetLabeled with high precision. If \adversary has a set $\datasetLabeled_{\adversary}$ of samples where it suspects that some of them are part of \datasetLabeled, it is sufficient if there is some overlap, thus $\datasetLabeled_{\adversary} \cap \datasetLabeled\not=\emptyset$. Given the widely available and well-known datasets, e.g., for image recognition and the fact that \adversary can try many samples, it is reasonable to assume that at least some of the guessed samples in $\datasetLabeled_{\adversary}$ are correct~(see \sect\ref{sect:eval-ablation}).\\
However, we do not make any assumptions about the training process. \adversary can only augment the unlabeled dataset but lacks knowledge of the used algorithms, DNN architecture, or training specifics. Neither can \adversary affect these parameters nor can \adversary monitor the training process.
Thus \adversary performs a black-box attack. Additionally, \adversary is unable to manipulate the labeled samples in \datasetLabeled, delete samples from \unlabeledDataset, or modify non-poisoned samples. The only action that \adversary can take is adding samples to \unlabeledDataset.

\noindent
Notably, we do not assume that the adversary \adversary has any insight into the active training process of the victim \victim. Consequently, \adversary cannot directly verify the success of the attack and must instead estimate its effectiveness through prior simulations, similar to existing work on adversarial deep learning~\cite{shan2024nightshade,shan2023glaze,liang2023adversarial}. This aspect will be further discussed in \sect\ref{sect:discussion-limitations}.

\subsection{Challenges}
\label{sect:problem-requ}
\noindent In SSL, it is not practical to make strong assumptions about the adversary \adversary. Instead, \adversary has to operate blindly and perform the attack, i.e., upload the images to a platform \platform, without knowing any details about the training (SSL algorithm, DNN architecture, training parameters, etc.). 
In contrast to existing work~\cite{franci2022influence,liu2019unified}, we consider an adversary \adversary that does not have these details. Instead, \adversary aims to negatively affect the performance of the trained model only by uploading samples to a publicly accessible platform that is designed for hosting user-generated content. This open threat scenario poses a number of significant challenges that we describe in the following:

\noindent\textbf{\challengeNew{1} - Prevent utilization of non-manipulated samples:} 

Given the assumption that \adversary cannot manipulate the samples uploaded by benign users to \platform but can only add additional samples to the platform, one of the main challenges is to manipulate a small number of samples in such a way that they negatively affect the ability of the training algorithm to benefit from the large number of remaining benign samples. As we show in \sect\ref{sect:eval-ablation}, simply preventing the manipulated samples from contributing positively to the training is not sufficient, as there are still enough benign samples for the DNN model to learn from.

\noindent\textbf{\challengeNew{2} - Stealthiness:} 
Even for small percentages of manipulated images, e.g., 10\%, users will notice if many media files on social networks show a certain pattern or artifacts that are caused by the poisoning. Additionally, if humans can detect these manipulations, it would be possible to gather a sufficient number of manipulated samples to develop an automated filtering system and thus defeat the attack. Therefore, another important challenge is to stealthily manipulate the samples to evade detection by humans while still effectively disturbing the SSL training algorithm.

\noindent\textbf{\challengeNew{3} - Causing SSL Algorithm to mislabel samples:}
Our attack makes use of wrongly labeled samples to disturb the training. However, the adversary \adversary does not have access to the labeled dataset and the labels for the unlabeled dataset are automatically determined during the training. Therefore, another challenge is to make the SSL algorithm to craft wrong labels for the manipulated samples without changing the sample (i.e., images) too much, as this would affect the ability of the DNN model to learn the wrong behavior.

\noindent\textbf{\challengeNew{4} - Effectiveness Against Benign Majority:} 
Given the vast amount of content posted daily on social media, it is unrealistic to assume that \adversary can control a majority of the unlabeled samples by uploading a large number of manipulated media files. Therefore, one of the key challenges for an untargeted poisoning attack is to effectively decrease the utility of a trained model by only manipulating a small fraction of the samples. For backdoor attacks, both, the benign objective of achieving a decent performance and the attacker's objective to inject a backdoor, can be fulfilled simultaneously. However, for untargeted poisoning attacks, the attacker's objective of preventing the model from achieving a decent performance is in contradiction with the benign objective. Only one objective can be achieved at the same time. Therefore, a key challenge is, how to succeed against the majority of benign samples by using only a minority of samples.

\section{\ourname}
\label{sect:approach}

\noindent In the following, we introduce the \ournameAttackNew. First, we describe the motivation behind the attack. Then, we first provide a high-level overview of the \ournameAttackNew (\sect\ref{sect:approach-design} followed by a detailed explanation of the individual components (\sect\ref{sect:approach-implementation}).

\begin{figure}[tb]
    \centering
	\begin{subfigure}[b]{0.12\textwidth}
         \centering
         \includegraphics[width=\textwidth]{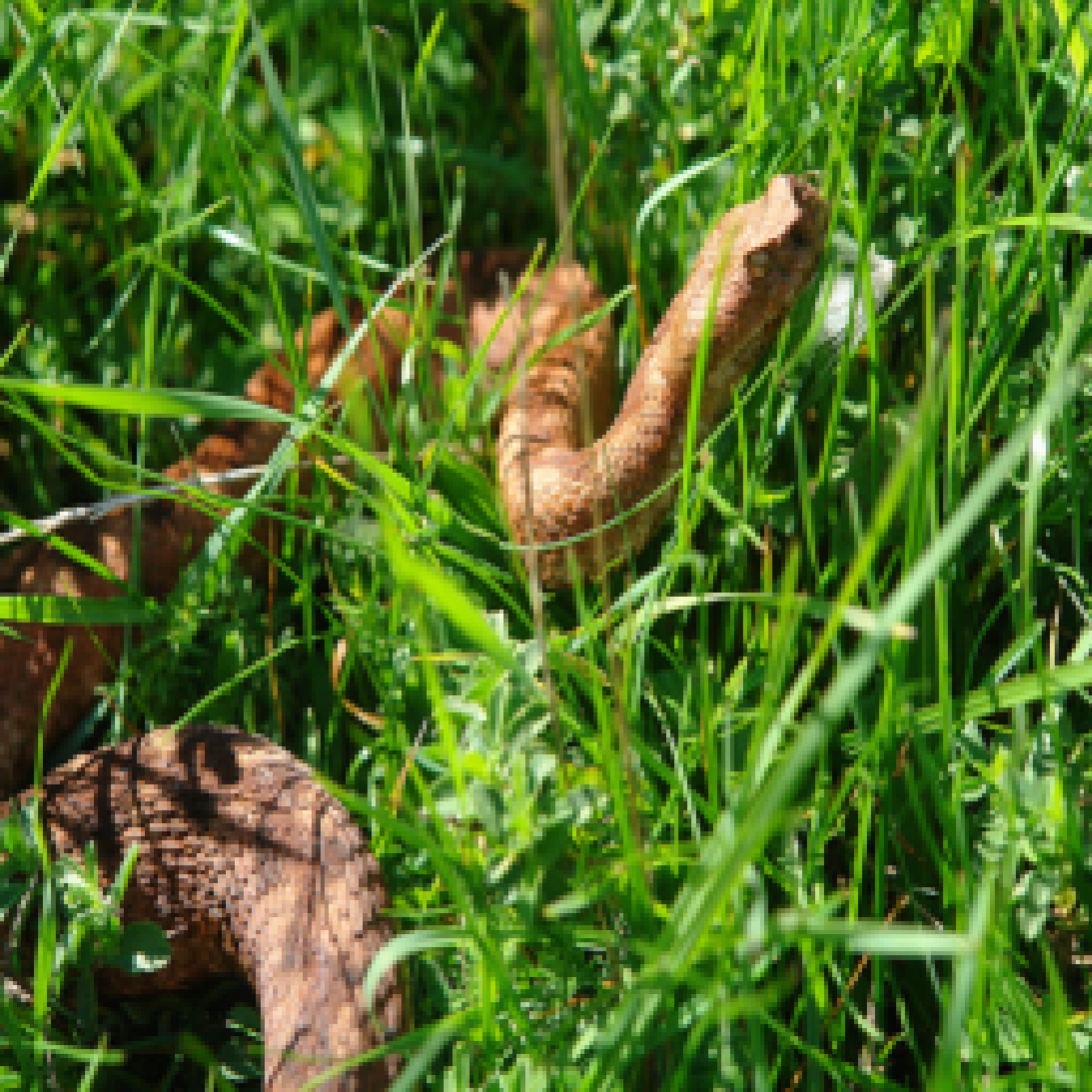}
         \caption{PV = $0\%$}
  	    \label{fig:pv:pv00}
     \end{subfigure}
     \hfill
      \begin{subfigure}[b]{0.12\textwidth}
         \centering
         \includegraphics[width=\textwidth]{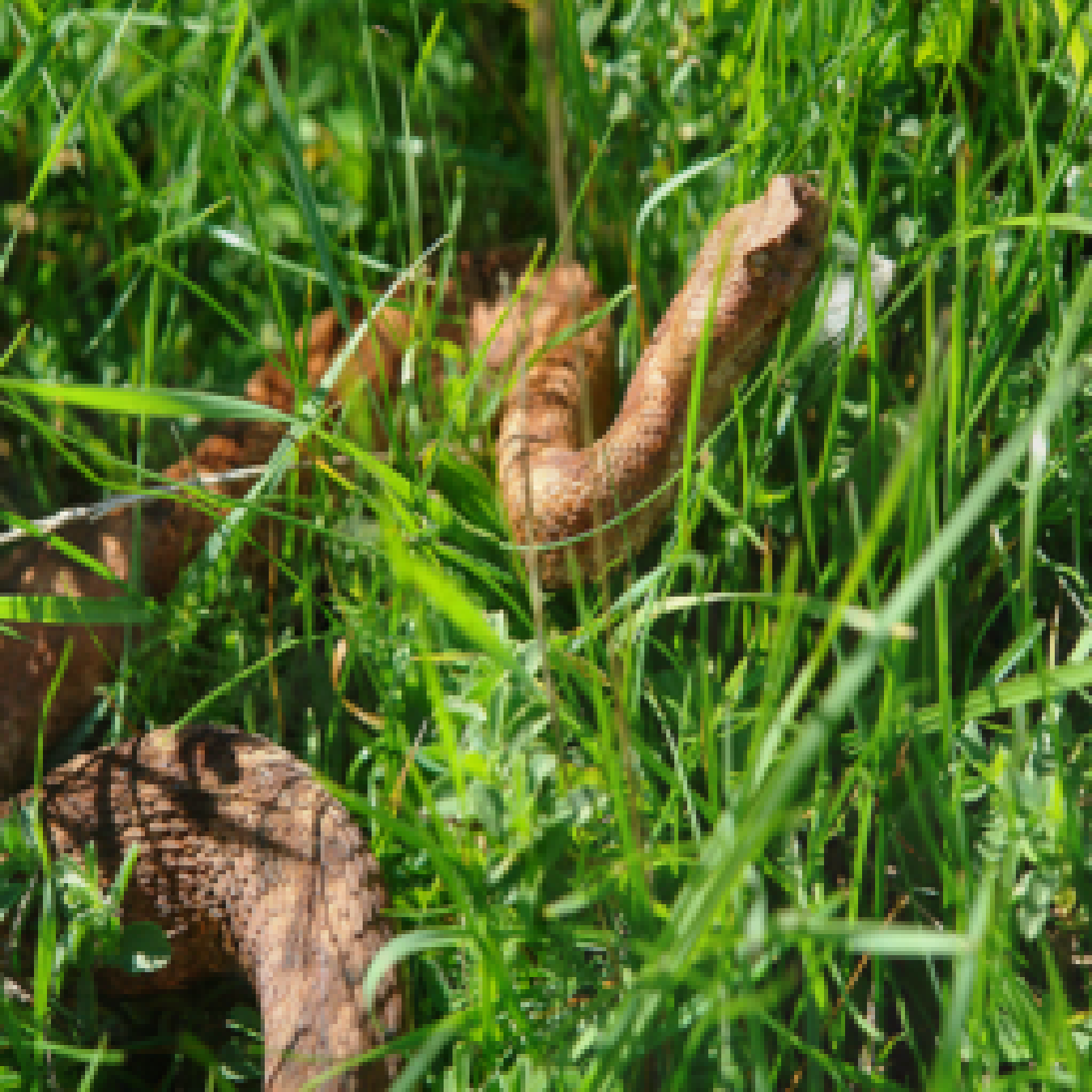}
          \caption{PV = $10\%$}
  	    \label{fig:pv:pv10}
     \end{subfigure}
     \hfill
      \begin{subfigure}[b]{0.12\textwidth}
         \centering
         \includegraphics[width=\textwidth]{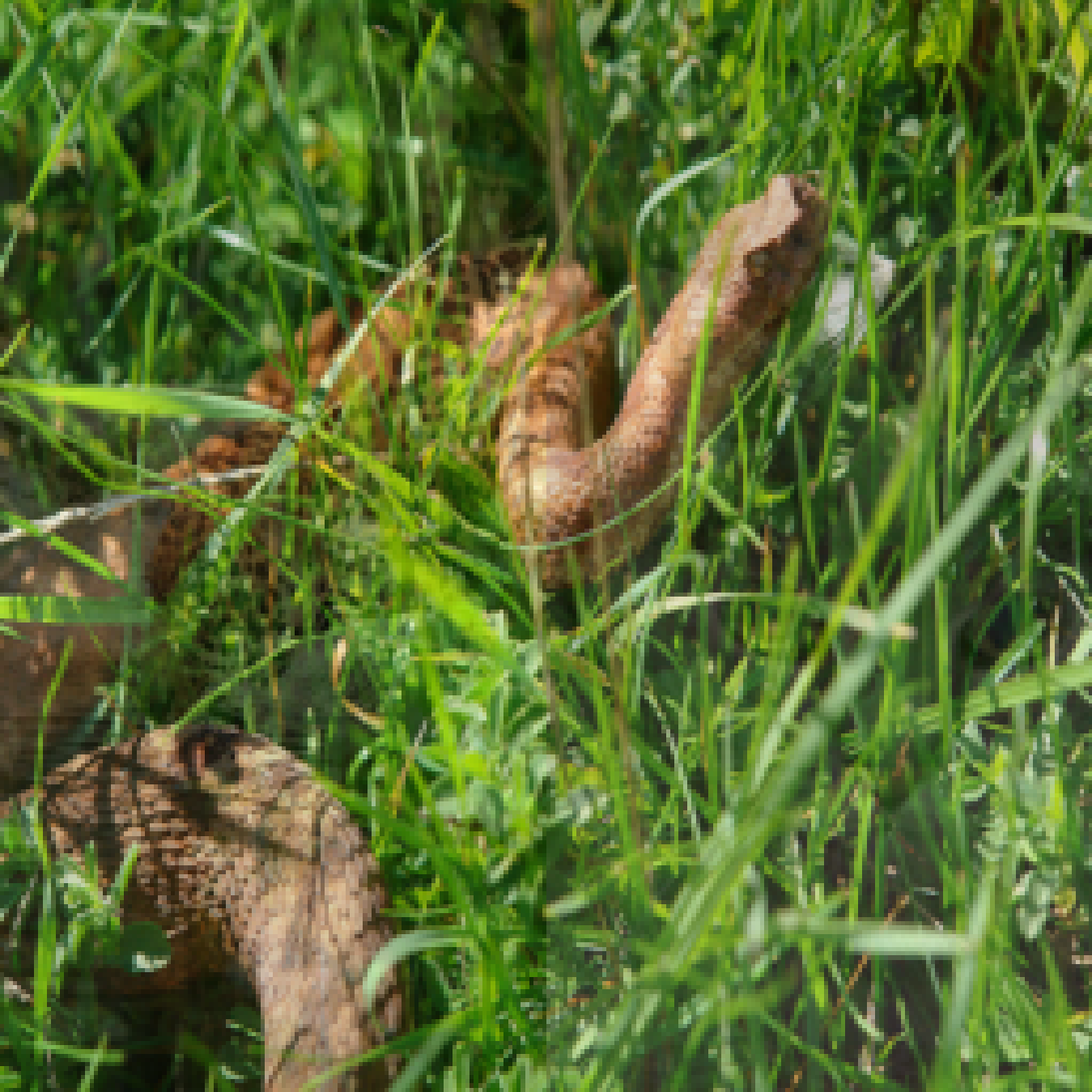}
          \caption{PV = $25\%$}
  	    \label{fig:pv:pv25}
     \end{subfigure}\\
     
	\begin{subfigure}[b]{0.12\textwidth}
         \centering
         \includegraphics[width=\textwidth]{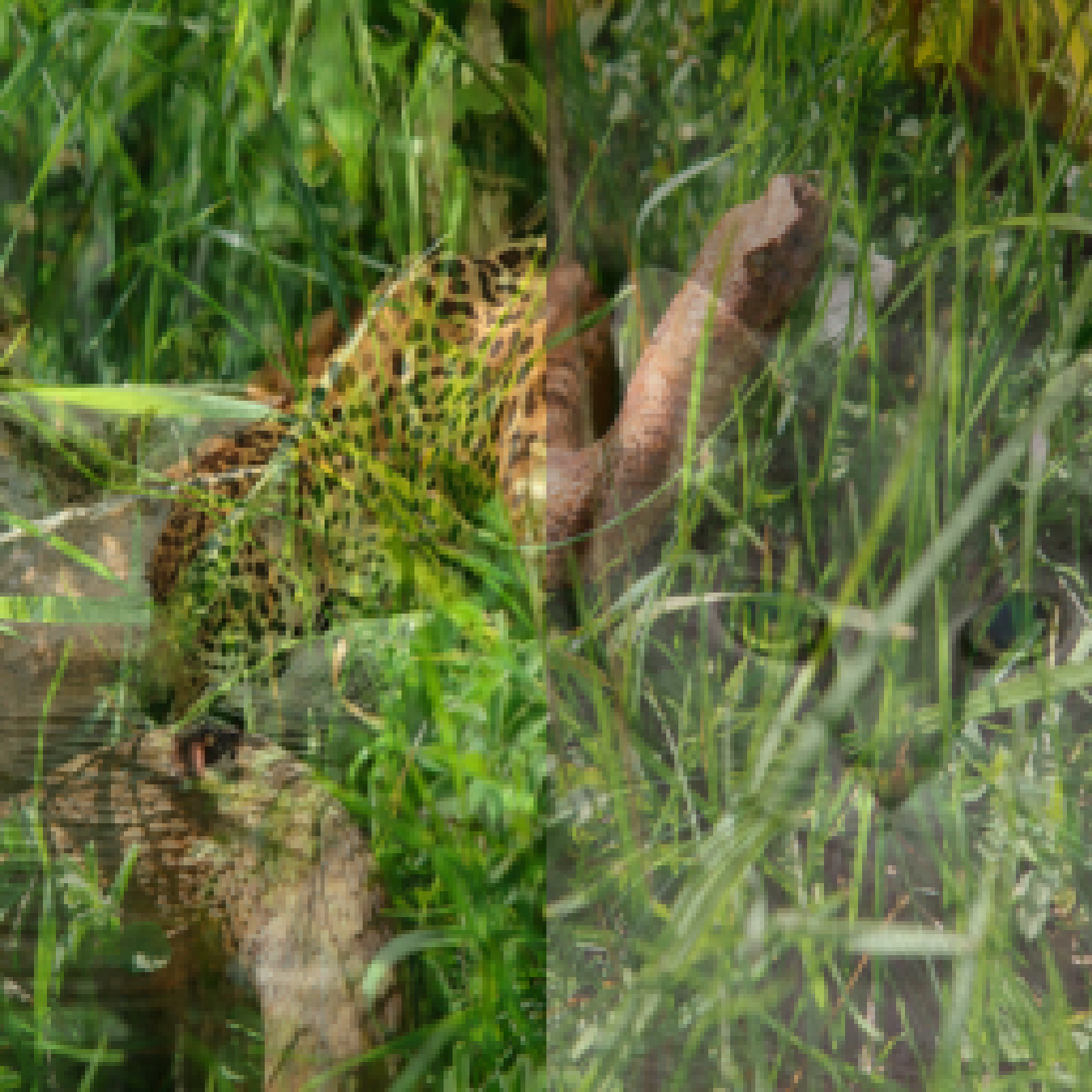}
         \caption{PV = $50\%$}
  	    \label{fig:pv:pv50}
     \end{subfigure}
     \hfill
      \begin{subfigure}[b]{0.12\textwidth}
         \centering
         \includegraphics[width=\textwidth]{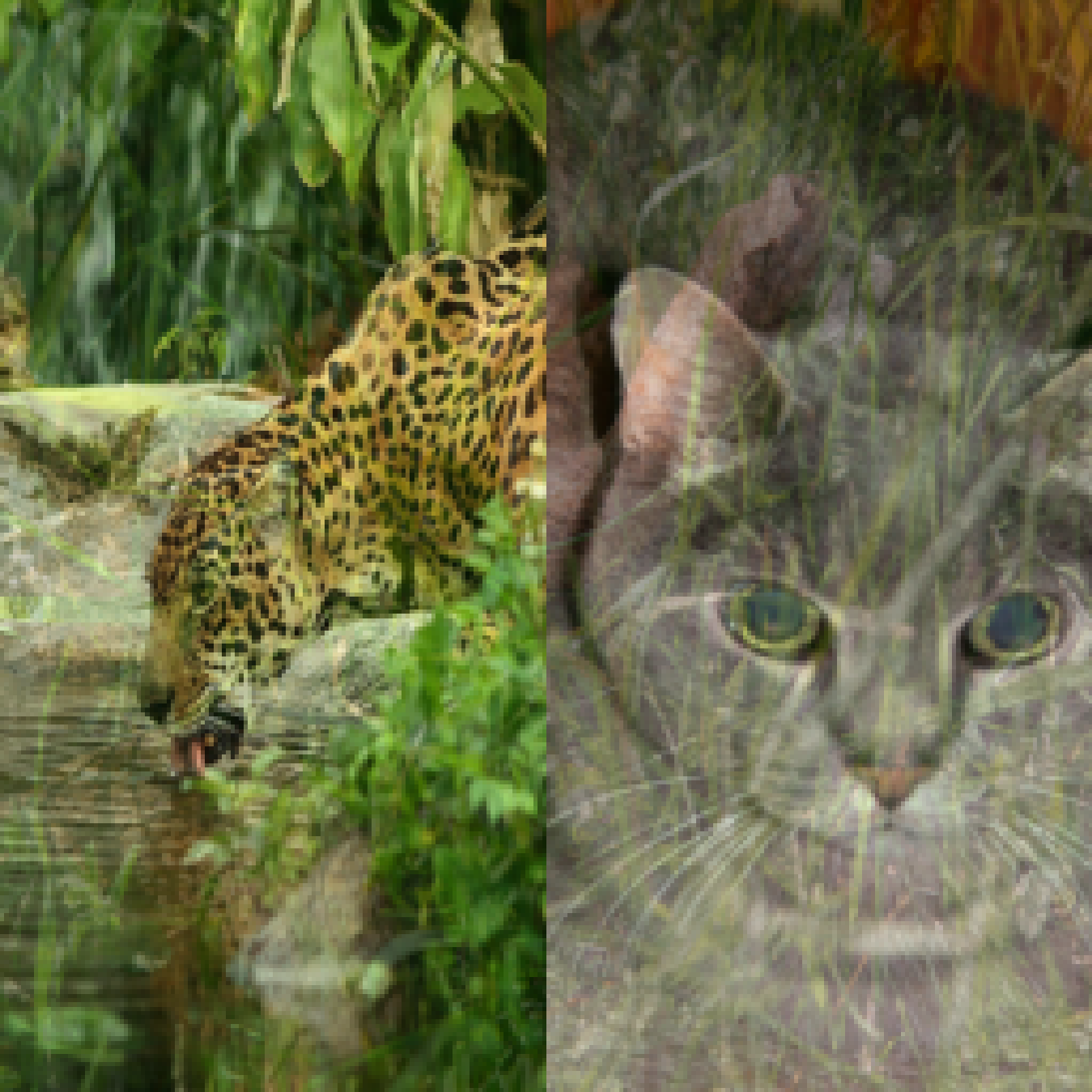}
          \caption{PV = $75\%$}
  	    \label{fig:pv:pv75}
     \end{subfigure}
     \hfill
      \begin{subfigure}[b]{0.12\textwidth}
         \centering
         \includegraphics[width=\textwidth]{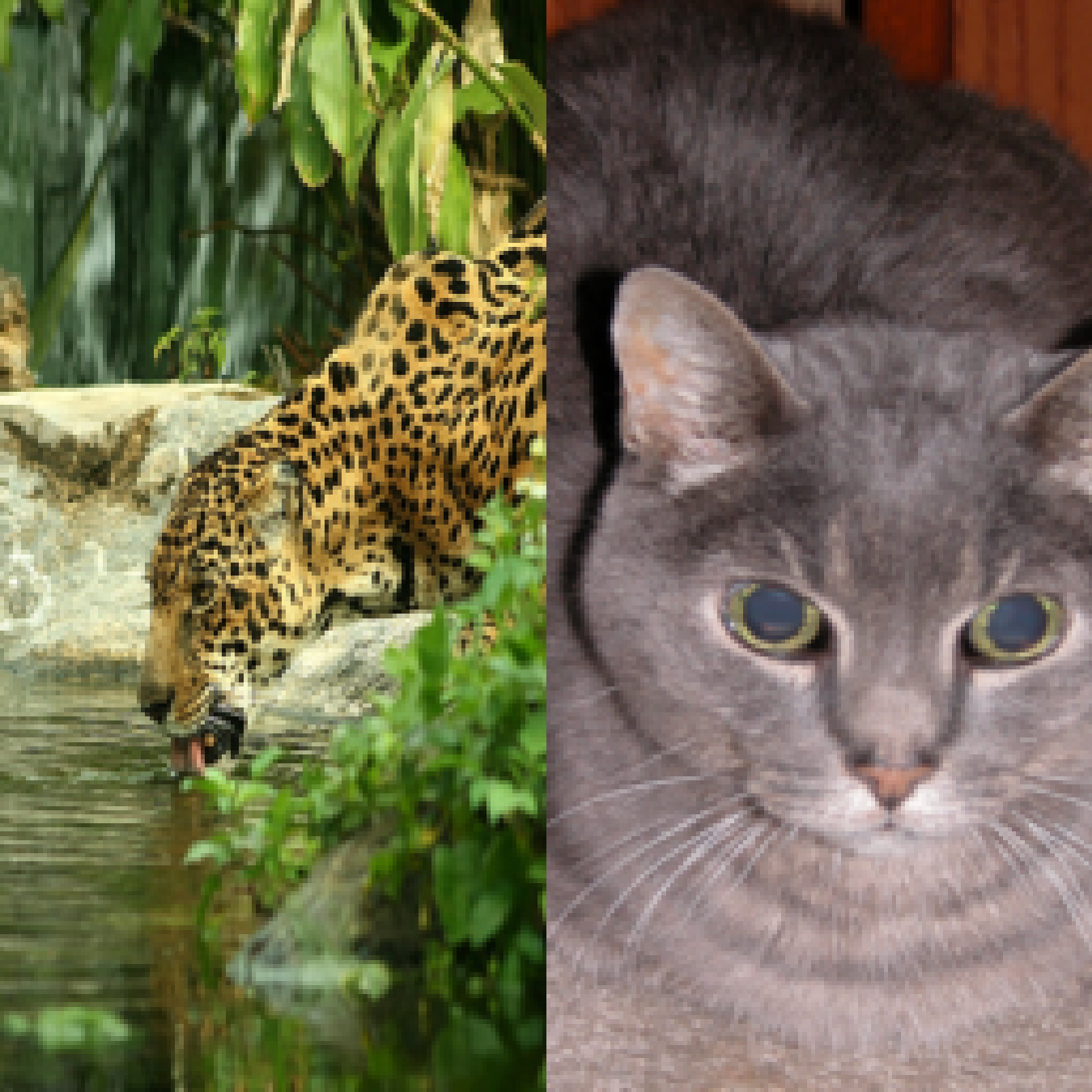}
          \caption{PV = $100\%$}
  	    \label{fig:pv:pv100}
     \end{subfigure}%\vspace{-0.5em}
     \caption{Effect of different Patter Visibility (PV) rates to obtain a manipulated image for the \imagenet dataset.}
        \label{fig:pv}%\vspace{-0.5em}
\end{figure}
\subsection{Motivation}
\label{sect:approach-motivation}

\noindent The primary motivation for \ournameGen mechanism stems from the observation that SSL algorithms tend to rely heavily on the labeled data during early training epochs. Already during the early epochs of the training, the DNN achieves high accuracy on the labeled dataset \datasetLabeled (see \appSect\ref{app:labeledEvaluation}). This dependency on \datasetLabeled makes the DNN model susceptible to overfitting on the samples in the labeled dataset \datasetLabeled, which can lead to misclassification of unseen data. This phenomenon becomes particularly relevant when an input $u \in \datasetUnlabeled$ contains artifacts that resemble an image from the labeled dataset \datasetLabeled with label $l$. In such cases, the model's increased sensitivity to detect even parts of labeled samples causes it to predict label $l$ for a new sample $u$, regardless of the actual class of $u$. Thus, the overfitting of the DNN results in a high sensitivity to recognizing labeled images, even when they are present only as weak shadows within the manipulated image. 

\noindent Another consequence of the DNN's overfitting on \datasetLabeled is its tendency to predict labels with high confidence for samples that are similar to those in \datasetLabeled. As a result, these similar samples are incorporated into the training process during early epochs, while the remaining samples are only gradually added and utilized later on.

\subsection{High-Level Overview of \ourname}
\label{sect:approach-design}

\noindent  The fundamental principle of the \ournameAttackNew is to induce the SSL algorithm to generate incorrect labels for samples in the unlabeled dataset. During training, the SSL algorithm's erroneous guesses for these labels cause the parameter optimizer to make incorrect adjustments to the DNN's parameters, thereby reducing the model's utility rather than improving it. To cause SSL algorithms to mislabel the unlabeled samples, \ourname injects a poisoning pattern into the original samples. Once the manipulated samples are crafted, they are uploaded to a platform with user-generated content, such as a social network. When the victim later crawls this platform for content and downloads media, such as images, being posted there, the manipulated samples are unknowingly incorporated into their unlabeled dataset.

\noindent The manipulated samples consist of a combination of the original sample and an individual poisoning pattern. This construction is exemplified for images in Fig.~\ref{fig:poisonedConstruction}. The resulting image comprises two parts or layers that are placed on top of each other, the original image and the shadow, referred to as the poisoning pattern.

\noindent \textbf{Causing Wrong Label Guesses.} To influence the label guesses of the SSL algorithm with the poisoning pattern, \ourname exploits the tendency of SSL algorithms to overfit on samples from the labeled dataset \datasetLabeled. This overfitting makes the model highly sensitive to recognizing samples from the labeled dataset, even when they are barely visible in the image. To leverage this effect, we construct the poisoning pattern using images that resemble those in the labeled dataset. This causes the SSL algorithm to use the barely visible poisoning pattern as the basis for its label prediction, addressing challenge \challengeNew{3}. By constructing the manipulated sample primarily from a regular sample rather than solely utilizing the poisoned pattern, the resulting sample appears also less suspicious and barely visible to human observers, as demonstrated in Fig.~\ref{fig:poisonedConstruction}. This approach addresses \challengeNew{2}.

\noindent \textbf{Effect on the Training Process.} Using a barely visible poisoning pattern and primarily the original image ensures that the original sample is mainly used for training. During the training process, the optimization algorithm that adapts the DNN's weights calculates the loss and gradients based on the entire image, including its original content. Since the original sample is more prominent, it is used to train the DNN model with the wrongly guessed label. The optimizer then adjusts every parameter of the DNN to increase the probability of the guessed (wrong) label. Thus, the parameters are also changed based on the image's primary content to maximize the probability of the current label.

\noindent As described in \sect\ref{sect:approach-motivation}, samples similar to those in the labeled dataset \datasetLabeled, such as those containing parts of them, are incorporated into the training process already during the early phases. This early integration of manipulated samples affects the predicted labels for benign data, affecting the algorithm's ability to benefit from these data, addressing \challengeNew{4}. The combination of the manipulated sample and the mistakenly guessed label for benign data contradicts the desired benign behavior, disrupting the training process and hindering the DNN model's ability to effectively utilize non-manipulated data, addressing \challengeNew{1}.

\noindent \textbf{Example.} Fig.~\ref{fig:poisonedConstruction} shows exemplary the process of crafting a poisoning image. The resulting image consists of an original image of a snake as well as a weak shadow of a cat and a tiger serving as the poisoning pattern. Since the superimposing of the poisoning pattern is only very weak, it is invisible to human eyes. However, the overfitting of the DNN model causes the DNN to focus on the cat and the tiger, such that this weak shadow is sufficient to classify this image as a cat (or tiger). Especially in the early phases, the DNN model is not capable of recognizing arbitrary objects, but rather it has learned to recognize specific images from the labeled dataset. As a result, the manipulated image is either labeled as a cat or as a tiger. However, once the wrong label is assigned to this image, the training process will consider the whole image. Therefore, the DNN will be trained to recognize the snake image as a cat/tiger, as the poisoning pattern is barely visible in comparison to the snake.

\subsection{Implementation}
\label{sect:approach-implementation}
\noindent Each manipulated sample $m$ is created as overlay from a regular sample $r$ and the poisoning pattern $p$. To construct the poisoning pattern in the image domain, \ourname selects for each sample $r$ two images $p_1, p_2$ with respective labels $l_1, l_2$ from the set $\datasetLabeled_\adversary$ of images that are suspected to be part of the labeled dataset \datasetLabeled. Notably, we leverage two samples $p_1, p_2 \in \datasetLabeled_{\adversary}$ for crafting the poisoning pattern. Besides causing further distraction, this also increases the probability that one of the guesses is actually part of \datasetLabeled. We noticed during our experiments that choosing the samples $p_1$ and $p_2$ such that they belong to different classes ($l_1\not=l_2$) increases the attack's effectiveness, as the ambiguous labeling causes further distraction. Additionally, targeting different classes increases the probability of triggering the model's overfitting, leading to incorrect label predictions for the current sample.

\noindent To obtain the poisoning pattern $p$ for image applications, $p_1$ and $p_2$ are first cropped at the left and right side of the image, ensuring their width is half that of $r$. Then, if the concatenation of these cropped images is denoted as $p$, the color $c$ of the pixel on position $x$, $y$ of the manipulated image $m$ is then given by:

\begin{equation}
    m_{x,y,c} = (1-\text{PV})\cdot r_{x,y,c} + \text{PV} \cdot p_{x,y,c}
\end{equation}

\noindent The Pattern Visibility (PV) parameter controls the ratio that combines the regular sample $r$ and the poisoned pattern $p$. Increasing the value makes it easier for the DNN model to spot the fractions of the labeled images, while a low PV value ensures that the manipulated images remain inconspicuous.

\noindent The impact of varying PV values on the manipulated image is shown in Fig.~\ref{fig:pv}, its impact on \ournameGen effectiveness is evaluated in \sect\ref{sect:eval-attackparams}. For a PV of 100\% (depicted in Fig.~\ref{fig:pv:pv100}) only the poisoning pattern is visible. As the PV value decreases from 100\% to lower values, the image appears increasingly less suspicious to the human eye, making it increasingly difficult to discern the pattern (as demonstrated in Fig.~\ref{fig:pv:pv10}). However, even small PV values are capable of causing a decline in the model's performance. For instance,  a PV value of 10\%, as seen in Fig.~\ref{fig:pv:pv10}, is sufficient to result in a 10\% decrease in the accuracy of the trained model (see \sect\ref{sect:eval-basic}).

\noindent A challenge faced by the adversary is obtaining knowledge about the actual labeled samples. As discussed in \sect\ref{sect:sota}, existing literature assumes unrestricted access to the labeled dataset, the ability to modify it, or even knowledge of the model's parameters, which may not be realistic. In the case of \ourname, \adversary is required to guess samples $\datasetLabeled_{\adversary}$ that are present in the labeled dataset \datasetLabeled. However, only some overlap between both is necessary, i.e., \mbox{$\datasetLabeled_{\adversary} \cap \datasetLabeled\not=\emptyset$.}

\noindent In order to incorporate two labeled samples into a single sample, we employ a strategy of cropping the central portion of the labeled samples. This approach has shown to be effective as the central region of an image typically contains the most salient features. It is worth noting, however, that \adversary can manually inspect the poisoned patterns before uploading the manipulated samples. As such, if \adversary determines that the most relevant part of a labeled sample is not located in the center, \adversary can easily adapt the cropping accordingly. Also, if \adversary considers some images to be too suspicious, e.g., if the poisoned pattern can be spotted in the image, it can replace the respective poisoned image with another version, e.g., by using a different benign image as the original image.

\section{Evaluation}
\label{sect:eval}
In previous sections, we introduced the \ournameAttackNew, which disrupts SSL training through the addition of manipulated samples to the unlabeled dataset. In the following, we show the effectiveness of \ourname on \nDatasets\xspace different datasets and conduct a real-world case study involving three prominent social media platforms, which are suitable data sources for SSL. In addition, we analyze in App.~\ref{app:gradcam} changes in the behavior of the model that are caused by \ourname using explainable AI, specifically saliency maps for benign and poisoned input samples.

\subsection{Experimental Setup}

\noindent \textbf{Datasets:} 
\begin{table}[tb]

    \caption{Overview of the used datasets}
    \label{tab:datasetsetup}

\begin{tabular}{l|rrrr}
    Dataset & \makecell{\#train\\samples} & \makecell{\#test\\samples} & \multicolumn{1}{c}{Model} & \makecell{\#DNN\\parameters}\\\hline
    \cifar & \numprint{50000} & \numprint{10000} & Wide ResNet-28-2 & \numprint{1469642} \\
    \mnist & \numprint{60000} & \numprint{10000} & Wide ResNet-28-2 & \numprint{1469354}\\
    \svhn & \numprint{604388} & \numprint{26032} & Wide ResNet-28-2 & \numprint{1469642} \\
    \gtsrb & \numprint{26640} & \numprint{12630} & Wide ResNet-28-2 & \numprint{1473899} \\
    \imagenet & \numprint{1281167} &\numprint{50000} & ResNet-50 & \numprint{25557032} \\
    \stl & \numprint{105000} & \numprint{8000} & Wide ResNet-37-2 & \numprint{5933770}\\
\end{tabular}
\end{table}

We employed \nDatasets different datasets that are regularly utilized to evaluate DNNs, particularly research that aims to address SSL from a security perspective~\cite{carlini2021poisoning,connor2022rethinking,feng2022unlabeled,franci2022influence,yan2021dehib,yan2021deep}. They are summarized in Tab.~\ref{tab:datasetsetup}.

\noindent\emph{\cifar}~\cite{krizhevsky2009learning} includes \numprint{50000} training images of 32$\times$32 pixels, featuring objects and animals from 10 distinct categories. The dataset is widely utilized as a benchmark for both SSL and DNN research~\cite{carlini2021poisoning,connor2022rethinking,feng2022unlabeled,franci2022influence,yan2021dehib,yan2021deep}.

\noindent\emph{\stl}\xspace is tailored explicitly for SSL and consists of \numprint{100000} unlabeled samples, \numprint{5000} labeled images, and \numprint{8000} test images. All images are colored and have a resolution of 96$\times$96 pixels~\cite{coates2011analysis}.
    
\noindent\emph{\svhn} comprises of \numprint{604388} training images and \numprint{26032} test images, with a resolution of 32$\times$32 pixels~\cite{netzer2011reading}.
     
\noindent\emph{\mnist} consists of \numprint{60000} training images showing handwritten digits, all of which are grayscale and having a resolution of 28$\times$28 pixels~\cite{lecun1998gradient}.

\noindent\emph{\gtsrb} (German Traffic Sign Benchmark) consists of \numprint{26640} train and \numprint{12630} test images of different traffic signs. Each of the 43 classes represents one traffic sign, while the images themselves show them at different daytimes, perspectives, and environments (urban and countryside)~\cite{houben2013detection}. The dataset was accessed via the official PyTorch integration~\cite{gtsrbTorch}.
   
\noindent\emph{\imagenet} is a large, high-resolution image dataset. We use the images from the ImageNet Large Scale Visual Recognition Challenge (ILSVRC), which consists of \numprint{1.3} million training images and \numprint{50000} test images from \numprint{1000} different categories. The size of the images is scaled to 256$\times$256 due to the large variety of image sizes~\cite{imagenet-object-localization-challenge}.\\

\noindent\textbf{Parameters:} 

The parameters used in this study align with the setup of Zhang \etal~\cite{zhang2021flexmatch}, where the models are trained for 256 epochs, which is sufficient for convergence. Only for \imagenet we used a larger number of epochs due to the more complex training task and larger number of trainable parameters. Since it is assumed that \adversary can only upload additional images to \platform but cannot affect the training process, no other parameters are changed from their default values. A cosine decayed learning rate is used, initialized to 0.03, a temperature $T$ of 0.5, and the loss weight for the unsupervised loss is set to 1.0 for \fixmatch and \uda, and 100.0 for \mixmatch. A confidence threshold of 0.8 is used for \uda and \mixmatch. The simulation of adding poisoned images to the unlabeled dataset is done by replacing a certain fraction of the unlabeled samples with manipulated images. The ratio of manipulated unlabeled samples to the total number of unlabeled samples is referred to as the Poisoned Data Rate (PDR).

\noindent To choose the number of labeled samples, we considered the number of classes and ensured that the benign setting achieved reasonable accuracy. 
For \cifar, the labeled dataset consisted of 40 images for \uda and \fixmatch. Only for \mixmatch 100 labeled examples were necessary to achieve a decent performance. For \stl we used 100 from \numprint{105000} for \uda and \fixmatch and 250 for \mixmatch. For \svhn, we utilized 40 labeled images out of \numprint{604388} images in total, from \gtsrb 129 from \numprint{26640}, for \mnist 10 out of \numprint{60000}, and for \imagenet \numprint{100000} out of \numprint{1281167} images. Aligned with established work on SSL~\cite{zhang2021flexmatch}, we randomly selected the labeled images.

\noindent\textbf{Experiment Environment:}

We evaluate our attack on three \sota SSL algorithms: \mixmatch, \uda, and \fixmatch as described in~\sect\ref{sect:background-ssl}. To provide a comprehensive overview of the performance of the \ournameAttackNew, we evaluate the overall performance (\sect\ref{sect:eval-basic}) and perform the case study (\sect\ref{sect:eval-casestudy}) for all three algorithms, while the evaluation of aspects that involve a large number of experiments, such as the performance for various PDR and PV rates, consider only one algorithm respectively. For the implementation, we used the NumPy~\cite{harris2020array} and PyTorch~\cite{pytorch} frameworks. Further, we used the SSL implementation of Zhang \etal~\cite{zhang2021flexmatch}. The experiments were conducted on three servers, one running Debian with 1 TB memory, an AMD EPYC 7742 CPU with 64 physical cores, and 4 NVIDIA Quadro RTX 8000, another server running Ubuntu with an Intel Xeon 5318S CPU having 24 cores, 512GB main memory, and 2 Nvidia RTX A6000. Due to the high computational effort for \imagenet, we used here another server with 2 AMD EPYC 7773x CPUs, 2 TB memory, and 3 NVIDIA H100 GPUs. To ensure consistent results, all experiments for \cifar, and \svhn were conducted on the first server, experiments for the \stl, \gtsrb, and \mnist datasets on the second server, and all \imagenet experiments were conducted on the third server, avoiding any minor inconsistencies due to different library versions.

\begin{table}[t]
    \caption{Basic Performance of \ourname for \cifar.}
    \label{tab:eval:basicPerformance}
    \centering
    \scaleTable{
    
        \begin{tabular}{l|rrr}
                        & \mixmatch & \uda & \fixmatch\\\hline
        Benign Scenario &  74.75\% & 79.64\% & 89.10\%\\
        Only Labeled Dataset & 32.23\% & 26.00\% & 26.00\% \\
        \hline
        Empty Images (PDR=50\%)& 25.52\% & 70.86\% & 74.04\%\\
        Removing 50\% samples & 63.25\% & 50.93\% & 75.17\%\\\hline
        \ourname (PDR=5\%, PV=0.1) & 64.85\% & 68.71\% & 83.68\%\\
        \ourname (PDR=50\%, PV=0.6) & 23.77\% &46.85\% & 36.11\%\\

        \end{tabular}
    }
\end{table}

\begin{table}[t]
    \caption{Basic Performance of \ourname for \stl.}
    \label{tab:eval:basicPerformance-stl}
    \centering
    \scaleTable{
    
        \begin{tabular}{l|rrr}
                        & \mixmatch & \uda & \fixmatch\\\hline
        Benign Scenario &  63.20\% & 78.90\% & 66.83\%\\
        Only Labeled Dataset & 39.53\% & 29.01\% & 29.01\%\\
        \hline
        Empty Images (PDR=50\%)& 38.10\% & 72.85\% & 64.63\%\\
        Removing 50\% samples & 60.91\%&  68.11\% & 60.03\%\\\hline
        \ourname (PDR=5\%, PV=0.1) & 60.34\% & 74.16\% & 61.63\%\\
        \ourname (PDR=50\%, PV=0.6) & 34.40\% & 49.16\% &44.33\%

        \end{tabular}
    }
\end{table}

\subsection{Performance of \ourname}
\label{sect:eval-basic}
\noindent The tables~\ref{tab:eval:basicPerformance} and \ref{tab:eval:basicPerformance-stl} show the effectiveness of the \ournameAttackNew on three \sota SSL algorithms (\mixmatch, \uda, and \fixmatch)  for \cifar and \stl. As Tab.~\ref{tab:eval:basicPerformance} demonstrates, a PDR of only 5\% is sufficient for \ourname to decrease the accuracy by~10\% for \mixmatch and \uda for \cifar and 4\% on average for \stl. With higher PDR values, the accuracy reduction becomes even more significant, with a drop to 23.77\% for \mixmatch, 46.85\% for \uda, and 36.11\% for \fixmatch. The table also includes two simple baselines. For one baseline (denoted as "Empty Images" in Tab.~\ref{tab:eval:basicPerformance}) half of the images are replaced by empty (black) images. For the other baseline (denoted as "Removing 50\% samples" in Tab.~\ref{tab:eval:basicPerformance}) the respective samples are removed. The results of these baselines demonstrate the superior effectiveness of the \ournameAttackNew. For example, for \fixmatch, using empty images or removing half of all images reduces the accuracy only to 74\% and 75\% respectively, while \ourname reduces the accuracy to 36\%, which is very close to the accuracy when only the labeled dataset is used (26\%). Further comparisons with baselines are provided in \sect\ref{sect:eval-ablation}.

\begin{table}[b]
    \caption{Effectiveness of \ourname for different datasets using \uda and a PDR of 5\% and PV of 0.1.}
    \label{tab:datasets}
    \centering
    \scaleTable{
    
        \begin{tabular}{l|rr}
        Dataset & Benign & Attack\\\hline
        \stl & 78.90\% & 74.16\%\\
        \mnist & 99.20\% & 92.97\%\\
        \svhn & 80.78\% & 64.87\%\\
        \gtsrb & 83.20\% & 77.78\%\\
        \imagenet\footnotemark & 68.08\% & 57.05\%\\
        \cifar & 79.64\% & 68.71\%\\
                
        \end{tabular}
    }
\end{table}

\noindent To assess the stability of the obtained results, we conducted multiple iterations of the experiments for each Semi-Supervised Learning (SSL) algorithm employing our method (\ourname) with varying seeds. We repeated the experiments 5 times with different seeds and compared the resulting accuracies. Notably, we found that the accuracies exhibited a standard deviation of 2.6\% on average across the different SSL algorithms, being negligible in comparison to the notable performance drop of almost 10\% induced by \ourname and underscoring its robustness.

\noindent Tab.~\ref{tab:datasets} shows the effectiveness of the \ournameAttackNew on various datasets that are commonly used as benchmarks for SSL algorithms and attacks. As shown, \ourname effectively undermines the learning process for all \nDatasets datasets, demonstrating its general applicability.

\subsection{Varying Attack Parameters}
\label{sect:eval-attackparams}

\begin{figure}[t]
    \centering
    \includegraphics[width=0.8\linewidth,trim={0 0.225cm -1cm .225cm},clip]{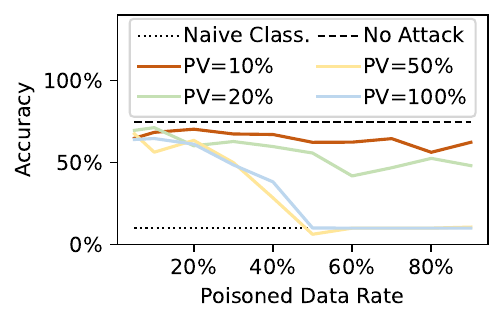}
    \caption{Impact of the Poisoned Data Rate (PDR) for the \ournameAttackNew for different Pattern Visibilities (PV) in comparison to the accuracy without attack (No Attack) and of a naive classifier (Naive Class.).}
    \label{fig:eval-pdr}
\end{figure}

\begin{figure}[t]
    \centering
    \includegraphics[width=0.8\linewidth,trim={0 0.225cm -1cm .225cm},clip]{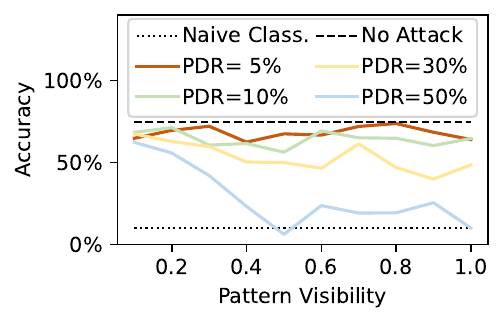}
    \caption{Impact of the Pattern Visibility for the \ournameAttackNew for different Poisoned Data Rates (PDR) in comparison to the accuracy without attack (No Attack) and of a naive classifier (Naive Class.).}
    \label{fig:eval-pv}
\end{figure}

\noindent The figures~\ref{fig:eval-pdr} and \ref{fig:eval-pv} show the effectiveness of the \ournameAttackNew for different PDRs and PVs using \mixmatch. As both figures illustrate, a PDR of 5\% and a PV of 0.10 suffice to reduce the performance by 10\% compared to the benign scenario. Both figures also show that the attack becomes more effective when using higher PDRs or higher PVs. As visible in Fig.~\ref{fig:eval-pdr}, by using a PDR of 50\%, the attack turns the trained model into a naive one, which always predicts the same class regardless of the input. It is worth noting that, if we use only the labeled dataset, the accuracy is still 32.23\%. Therefore, without providing any incorrect labels but causing the SSL algorithm to mislabel the samples itself, the \ournameAttackNew causes the resulting model to perform worse than not using any unlabeled sample.

\footnotetext{Due to the complex task and large number of classes, for \imagenet we use the top-5 accuracy. Thus, a sample is considered to be predicted correctly, if the true class is among the five classes with the highest predicted probabilities. Therefore, \ourname must not only make sure that the correct class does not receive the highest probability but that it is not even among the 5 classes with the highest predicted probabilities.}

\subsection{Ablation Study}
\label{sect:eval-ablation}

\noindent To the best of our knowledge, \ourname is the first untargeted poisoning attack that does not require any control over the victim. Therefore, no similar attacks exist for comparison. To evaluate the advantage of \ourname, we therefore defined a number of baselines and also adapted two backdoor attacks~\cite{nguyen2021wanet,carlini2021poisoning} to utilize them for an untargeted poisoning attack. Table~\ref{tab:ablationstudy} presents the performance of various alternative options for the \ournameAttackNew, in which 10\% of the data can be manipulated or removed. As shown in the table, the baseline performance of \uda, without any attack, is 79.64\% accuracy. However, when utilizing only the labeled dataset, the accuracy drops to 26.00\%.

\noindent For two straightforward baseline attacks, we prevent the SSL training from taking any advantage from the manipulated samples. A naive strategy would be simply not to upload these samples to the platform with user-generated content, to reduce the size of the unlabeled dataset. In this case, only the benign samples can be utilized for training. A second straightforward baseline attack to prevent the training from utilizing the manipulated images is uploading empty images, e.g., showing only black pixels. However, the results in Tab.~\ref{tab:ablationstudy} show that only preventing the model from utilizing the manipulated examples, therefore, not uploading them, is not an effective method for undermining the model. % as the overfitting cannot be exploited.
Additionally, using empty (black) images as an alternative option also demonstrates no significant impact on performance. However, it is worth noting that this strategy seems to have a regularization effect on the training process, thereby focusing the guessed probability distribution on the correct label and thus improving the model's accuracy. These experiments, therefore, showed that it is not sufficient to prevent the utilization of the manipulated samples but the utilization of the non-manipulated samples must be prevented for a successful attack (see challenge~\challengeNew{1}).

\noindent In recent years, several backdoor attacks against SSL have been developed that involve poisoning the unlabeled dataset. For a more sophisticated baseline, we adapted these approaches to make the model overfit and thus reduce the model's performance. The original backdoor attacks utilize interpolation to establish a link between samples that are intended to be mislabeled and the target samples~\cite{carlini2021poisoning,connor2022rethinking}. 
To adapt this attack and perform an untargeted poisoning attack, we inserted a colored shape on the images, causing the DNN to overfit and focus solely on the presence of this pattern. We then utilize interpolation to create a bridge between this pattern and samples from other classes. The rationale is that the model overfits only on the pattern and unlearns other class-specific properties. Aligned with the results of Carlini~\cite{carlini2021poisoning}, we use the density function $p(x)= 1.5 - x$ for interpolation. Results using other density functions are provided in App.~\ref{app:density}.

\noindent 
We evaluated various baseline attacks and compared their effectiveness against \ourname. First, we implemented Carlini's backdooring approach~\cite{carlini2021poisoning}, which interpolates between samples of different classes to misclassify samples, effectively creating a bridge between the two classes (referred to as Sample-Interpolation in Table~\ref{tab:ablationstudy}). The rationale is that this interpolation might cause the SSL algorithm to incorrectly label samples, hindering its ability to accurately learn the characteristics of each class.
Additionally to the attack of Carlini, we adapted several attack strategies that were originally developed for centralized learning to SSL. This included a sophisticated backdoor trigger injection technique using image warping~\cite{nguyen2021wanet} and also an adversarial-example-based attack employing the Fast Gradient Sign Method (FGSM) proposed by Goodfellow \etal~\cite{goodfellow2014explaining}. However, as shown in Tab.~\ref{tab:ablationstudy}, all three attacks were ineffective. Specifically, the FGSM attack reduced the accuracy only to 76.86\% (PDR=5\%) or 74.34\% (PDR=10\%), while \ourname reduced the accuracy to 68.71\% (PDR=5\%) and 67.15\% (PDR=10\%). The backdoor-inspired attacks failed, most likely due to the fact that in an untargeted poisoning attack, it is not sufficient to add a bridge to the data for making a few, well-defined samples misclassified. For the FGSM attack, the inability to prevent the utilization of the benign samples, given that the attacker controls only a small fraction of the unlabeled dataset \unlabeledDataset, contributed to its failure. The attack needs to prevent the utilization of benign data to succeed (see \challengeNew{1}), since the attack objective (preventing high accuracy) contradicts the benign objective (achieving high accuracy), which is given by the majority of the data. Simply connecting a few samples from specific classes or preventing the utilization of a few samples is, therefore, insufficient. In comparison, \ourname exploits the model's overfitting on the labeled dataset \datasetLabeled and causes the model to mislabel the unlabeled samples.

\noindent Also the baseline of removing the manipulated images from the dataset has a negligible impact on performance. A reason might be that, even after removal, still enough samples remain that can be used for training the DNN. This emphasizes that it is not sufficient to prevent the poisoned samples from being used it is necessary that the manipulated samples affect the ability to utilize the other samples in the unlabeled datasets, which were not manipulated.

\begin{table}[t]
    \caption{Effectiveness of different variations of \ourname for a PDR of 10\%.}
    \label{tab:ablationstudy}
    \centering
    \scaleTable{
    
        \begin{tabular}{l|rrr}
        Attack & \uda\\\hline
        Benign Scenario &  79.64\% \\
        Use only Labeled Dataset & 26.00\%\\
        \hline
        Two-unlabeled attack  & 79.92\%  \\
        Empty Images & 89.06\%\\
        Remove 10\% Samples & 77.01\%\\
        Sample-Interpolation~\cite{carlini2021poisoning} & 81.03\%\\
        Warp Trigger~\cite{nguyen2021wanet} & 78.11\%\\
        FGSM~\cite{goodfellow2014explaining} & 74.34\%\\
        Three-Unlabeled-One-Labeled & 71.72\%\\\hline
        \ourname (PV=0.2, PDR=10\%) & 67.15\% \\
        \ourname (PV=0.1, PDR=5\%) & 68.71\% \\
        \end{tabular}
    }
\end{table}
\noindent In comparison to these baselines, \ourname reduces the performance by more than 12\% compared to the baseline performance, demonstrating its effectiveness in not only rendering the manipulated samples unusable but also preventing the training algorithm from utilizing a significant portion of the non-manipulated images.

\noindent The table also shows the accuracy when the attacker utilizes a pattern that is composed of 4 images, with only one of them being a part of the labeled dataset. This scenario is particularly relevant in situations when the attacker must infer the labeled samples based on typical datasets for the targeted application. As the table shows, this attack is still effective and causes a significant drop in the accuracy.
In addition, we performed an experiment where the set $\datasetLabeled_{\adversary}$, containing the samples that the adversary believes to be used as labeled samples, overlaps only by 20\% with the set \datasetLabeled of the actual labeled samples. Here, \ourname still reduced the model's utility to 71.38\%.

\begin{table}[t]
    \caption{Performance of \ourname for \cifar if 10\% of the labeled data are guessed correctly (precision=10\% and sensitivity=10\%) using PDR=5\% and PV=0.1.}
    \label{tab:eval:lowOverlapp}
    \centering
    \scaleTable{
    
        \begin{tabular}{l|rrr}
                        & \mixmatch & \uda & \fixmatch\\\hline
        Benign Scenario &  74.75\% & 79.64\% & 89.10\%\\
        
        \hline
        \ourname & 64.85\% & 68.71\% & 83.68\%\\
        \ourname Reduced Overlap & 70.04\% & 73.71\% & 84.60\%

        \end{tabular}
    }
\end{table}

In addition, we evaluated also scenarios, where the attacker's guesses about the labeled dataset \datasetLabeled are made with low precision and sensitivity. Particularly, only 10\% of the suspected samples $\datasetLabeled_{\adversary}$ are actually contained in \datasetLabeled (precision=10\%), while the correctly guessed samples make only 10\% of the actually labeled dataset \datasetLabeled (sensitivity=10\%). As Tab.~\ref{tab:eval:lowOverlapp} shows, although with reduced ability to guess the labeled samples, \ourname remains effective, as the accuracy is still reduced by approx. 5\%.

\subsection{Potential Countermeasures}
\label{sect:eval-aeDefenses}
\begin{figure}[tb]
    \centering
	\begin{subfigure}[b]{0.4\textwidth}
         \centering
         \includegraphics[width=\textwidth,trim={0.05cm 0.3cm -0.6cm 0cm},clip]{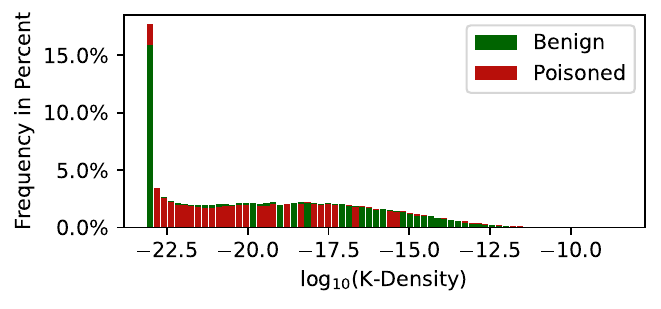}
         \caption{PV = 10\%}
  	    \label{fig:thresholding:pv010}
     \end{subfigure}
	\begin{subfigure}[b]{0.4\textwidth}
         \centering
         \includegraphics[width=\textwidth,trim={0.05cm 0.3cm -0.6cm 0cm},clip]{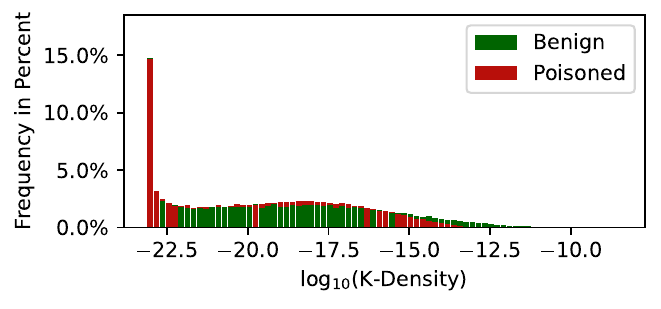}
         \caption{PV = 50\%}
  	    \label{fig:thresholding:pv050}
     \end{subfigure}
	\begin{subfigure}[b]{0.4\textwidth}
         \centering
         \includegraphics[width=\textwidth,trim={0.05cm 0.3cm -0.6cm 0cm},clip]{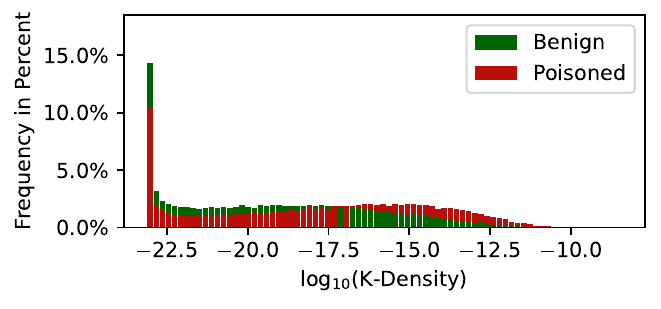}
         \caption{PV = 75\%}
  	    \label{fig:thresholding:pv075}
     \end{subfigure}

	\begin{subfigure}[b]{0.4\textwidth}
         \centering
         \includegraphics[width=\textwidth,trim={0.05cm 0.3cm -0.6cm 0cm},clip]{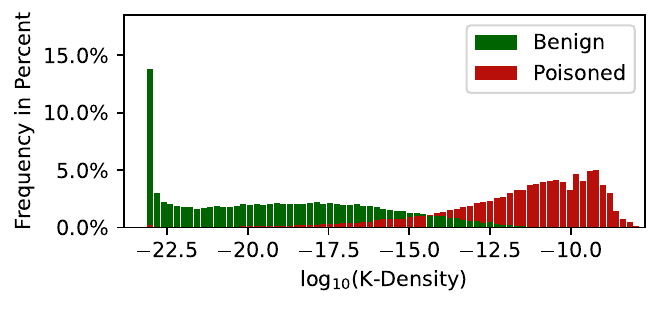}
         \caption{PV = 100\%}
  	    \label{fig:thresholding:pv100}
     \end{subfigure}
     
     \caption{Distribution of kernel density scores (K-Density)~\cite{pang2018towards} for benign and poisoned images using different Pattern Visibilities (PVs).}
        \label{fig:defense-thresholding}
\end{figure}

The poisoning pattern of \ourname shows some similarities to adversarial examples (see \sect\ref{sect:discussion-defenses}). In the following, we evaluate \ournameGen effectiveness in the presence of potential countermeasures. Zantedeschi \etal proposed applying Gaussian noise on the inputs~\cite{zantedeschi2017efficient}. However, as shown in Tab.~\ref{tab:defenses}, the noise does not mitigate the attack. Pang \etal\xspace proposed analyzing the states of the final hidden layer for a sample $x$ to determine if $x$ contains an adversarial pattern~\cite{pang2018towards}. Given the predicted label $y$, all training samples $X_y$ having the label $y$, $f_z$ the current model until the final hidden layer, and the Gaussian Kernel $k(\cdot, \cdot)$, then the K-density score $KD(x)$ for $x$ is defined as:
\begin{equation}
    KD(x) = \frac{1}{|X_y|}\cdot\sum\limits_{x_i\in X_y}k\left(f_z\left(x\right), f_z(x_i)\right)
\end{equation}
Since the thresholding test was developed for scenarios where fully labeled data are available, we used the softmax label of the training data to craft $X_y$. Fig.~\ref{fig:defense-thresholding} shows the distribution of scores for benign and poisoned images using a PDR of 5\% and different PV values after the model was trained for 10 epochs. As the figure shows, the distribution of scores for benign and manipulated samples are indistinguishable for PV $\leq$ 50\%, and only for larger PVs could the metric identify some values. Notably, for such a high PV value, humans can already identify the manipulated images easily (see Fig.~\ref{fig:pv}).

\noindent We also trained a dedicated CNN classifier using the VGG-11 architecture. We trained the classifier on benign and poisoned \cifar samples and cross-evaluated them on benign and poisoned \imagenet samples (1.2M samples each). We measured the ratio of manipulated samples that are detected (True Positive Rate, TPR), the ratio of manipulated samples compared to the total number of samples that are classified as manipulated (Precision, PRC), F1-Score, and the overall ratio of correctly classified samples (Accuracy, ACC). As Tab.~\ref{tab:defenseclassifier} shows, the classifier was barely able to recognize samples for low PV values (TPR=44.52\%) and only for large PV values the classifier is effective. Notably, for PVs that are too large, the effectiveness is reduced again, showing that the classifier learned to detect the sample overlay.

\begin{table}[t]
    \caption{Effectiveness of DNN Classifier in recognizing \ourname attack in terms of True Negative Rate (TNR), True Positive Rate (TPR), Precision (PRC), F1-Score, and Accuracy (ACC) for different Pattern Visibilities (PV).}
    \label{tab:defenseclassifier}
    \centering
    \scaleTable{
    
        \begin{tabular}{r|rrrr}
        PV & \multicolumn{1}{c}{TPR} & \multicolumn{1}{c}{PRC} & F1-Score & \multicolumn{1}{c}{ACC}\\\hline
        10\% & 44.52\% & 73.62\% & 55.49\% & 64.28\%\\
        20\% & 68.94\% & 81.20\% & 74.57\% & 76.49\%\\
        50\% & 86.23\% & 84.39\% & 85.30\% & 85.14\%\\
        75\% & 73.82\% & 82.23\% & 77.80\% & 78.93\%\\
        100\% & 8.90\% & 35.81\% & 14.26\% & 46.47\%
        
        \end{tabular}
    }
\end{table}

\subsection{Case-Study}
\label{sect:eval-casestudy}

\noindent In addition to the aforementioned evaluation, we also conducted a case study to assess the effectiveness of our attack in a real-world scenario. Specifically, we uploaded manipulated images to three real-world social media platforms (Facebook, Instagram, and Pinterest) and evaluated how the platforms' image processing pipelines affect \ournameGen performance. To ensure that our experiments do not negatively impact other parties, we ensured that the images were only accessible to us by setting the privacy settings accordingly (see App.~\ref{app:eval-ethics}). Due to the limited number of images that can be uploaded\footnote{Notably, as described in \sect\ref{sect:discussion-impact}, an attacker could also use a script to upload each image automatically. However, as the usage of bots is clearly forbidden by these platforms, for our evaluation, we chose the batch strategy instead.} to Instagram and Pinterest, we stacked multiple images to form a larger image with a higher resolution, which was then split again after uploading and downloading. Since Instagram only allowed squared images, we padded the images using black pixels and removed the padding after the download. Additionally, Instagram and Pinterest might change the image scaling if it differs from the desired sizes. Therefore, we used a size of 612$\times$612 pixels for our images, as this size remained consistent.

\noindent As Tab.~\ref{tab:casestudy} shows, the \ournameAttackNew is effective for the data that were obtained from all three social networks. \ourname reduces the accuracy of the trained model by 10\%. Only in two cases, for \uda the accuracy is reduced only by 5\% (Facebook) or 7\% (Instagram). On the other side, for Instagram using \fixmatch, the accuracy is reduced by more than 25\%. Therefore, the augmentation, such as compression, that these social networks apply does not significantly affect \ournameGen impact.\\

\begin{table}[t]
    \caption{Effectiveness of \ourname in terms of accuracy for the \cifar dataset after downloading attack samples from different real social networks and comparison baselines.}
    \label{tab:casestudy}
    \centering
    \scaleTable{
    
        \begin{tabular}{l|rrr}
        Scenario & FixMatch & MixMatch & UDA\\\hline
        Benign Scenario & 89.10\% & 74.75\% & 79.64\% \\
        Sample Removal  & 84.58\% & 72.97\% & 77.01\% \\
        Simulation  & 77.07\% & 68.22\% & 67.15\% \\
        \hline
        Instagram  & 63.30\% & 63.86\% & 72.54\%\\
        Facebook  & 76.44\% & 63.12\% & 74.67\%\\
        Pinterest  & 72.90\% & 61.85\% & 69.65\%\\

        \end{tabular}
    }
\end{table}

\subsection{Partial Knowledge on Output Space}
\label{app:eval-output}
Recently, Richards \etal described the challenge of partial knowledge of the output space~\cite{richards2021adversarial}. For crafting the poisoning pattern, it is important that the patterns' labels differ from the ground truth label of the original image. To evaluate the effectiveness of \ourname under limited knowledge of the output space, we tested \ourname in two scenarios where the individual classes of \cifar were grouped into three actual classes. These groups included mechanical classes (airplanes, automobiles, ships, trucks), small animals (birds, cats, frogs), and large animals (deer, dogs, horses). 

\noindent In the first scenario, \adversary is aware of only these three classes, while the \victim uses all 10 classes. In the second scenario, \adversary uses all 10 classes, while the \victim uses only three classes. Since we observed a variance of 5\% in the experiment results, each experiment was repeated 10 times. We observed that reducing the knowledge of the output space affects \ournameGen effectiveness. In the first scenario, \ourname reduced the accuracy in average by 4.4\%, demonstrating that \ourname remains effective in this scenario. Notably, in the second scenario, no accuracy drop was observed, and \ourname was unable to reduce the accuracy.

\noindent This is likely because \ourname crafts the poisoning pattern using images with labels different from the original image, which causes the SSL algorithm to mislabel the images. However, if \adversary assumes a larger output space than \victim actually uses, it is probable that the images of the poisoning pattern belong to the same class. For instance, if \adversary creates a poisoning pattern for an airplane using images of ships and trucks, and \victim uses a reduced output space where these classes are grouped together into a superclass "mechanical items," the image will not be mislabeled during training. We will discuss this limitation further in \sect\ref{sect:discussion-limitations}.\\

\noindent In this section, we conducted a comprehensive evaluation of the effectiveness of the \ournameAttackNew using \nDatasets diverse datasets, showing its efficacy in distracting the training process diminishing the utility of the trained model, even for small PDRs. We evaluated the effects of various image augmentation techniques and potential counter measures, demonstrating the robustness of the \ournameAttackNew In a case study on 3 real-world social media platforms, we showed their susceptibility to \ourname, effectively demonstrating its effectiveness and the associated risks it entails.
\section{Security Considerations}

\noindent In the preceding sections, we introduced the \ournameAttackNew (\sect\ref{sect:approach}) and evaluated it in various scenarios (\sect\ref{sect:eval}). In this section, we will discuss the potential risks that the \ournameAttackNew presents (\sect\ref{sect:discussion-impact}), its limitations (\sect\ref{sect:discussion-limitations}), and possible future research direction (\sect\ref{sect:discussion-future}).

\subsection{Impact of the \ourname Attack}
\label{sect:discussion-impact}

\noindent The attack allows the adversary to degrade the performance of the model by simply uploading manipulated images to a platform with user-generated content. The poisoned pattern introduced by the attack confounds the training algorithm, causing it to mislabel the manipulated images and resulting in the DNN model being trained with incorrect data. Thus, by uploading few manipulated samples, the \ournameAttackNew affects not just a single model but all models being trained on these data.

\noindent Notably, the \ournameAttackNew does not have strong requirements or assumptions. It is successful even with small Pattern Visibilities (PV), making it difficult to detect the manipulated images. The \ournameAttackNew operates without any knowledge of the SSL algorithm, hyperparameters, or DNN architecture. Instead, the attacker simply uploads manipulated images to a platform with user-generated content, prior to the victim downloading the data and beginning the training process. This ability to poison the training without any knowledge of the attacked system highlights the increased attack surface created by utilizing data from untrusted sources without proper mitigation strategies. By uploading manipulated data, the attacker not only disrupts a specific SSL training procedure but also has the potential to prevent any SSL training for a particular task on the data from the affected platform. It is already sufficient to poison a small fraction of the data to achieve an accuracy drop between 5\% and 10\%. Notably, the uploading process could be also automated, as described in \sect\ref{sect:discussion-limitations}.

\noindent Even such small decreases in accuracy, such as a drop of 5\% or 10\% compared to the non-attacked model, can have serious consequences. For instance, in the context of self-driving cars, if a model fails to recognize the correct object, even in just 1\% of the cases, this will cause the car to show wrong and potentially dangerous behavior in such situations, making the vehicle unusable. Similarly, in competitive scenarios, if two actors both train models but one model performs 10\% better than the other model, the actor with the more accurate model will have an advantage in selling products based on their DNN models.

\noindent Also, the attack is not restricted to the image domain. The only requirement is the ability to combine a labeled sample with other samples, therefore, to overlay different samples weighted. This allows to combine the original samples with poisoning patterns. In \sect\ref{sect:approach-implementation}, we described this exemplary for images. However, this can also be applied in other applications, such as applications that process audio files. Also here, different samples can be overlayed and the volume of the different samples can be used as Pattern Visibility (PV). Therefore, the \ournameAttackNew is not restricted to image applications but is generally applicable. 

\noindent Thus, the combination of low requirements and the ability to reduce the accuracy, the \ournameAttackNew demonstrates the risk of untargeted poisoning attacks on SSL.

\subsection{Limitations of the \ourname Attack}
\label{sect:discussion-limitations}

\noindent The \ournameAttackNew does not require any prior knowledge of the training algorithm or the hyperparameters used, but it does necessitate some knowledge of the samples utilized for the labeled dataset. Adversaries can make educated guesses on the used labeled samples based on commonly available and publicly accessible datasets such as tiny images. In scenarios targeting a specific victim \victim, the attacker could also incorporate knowledge obtained through data breaches or insider threats, which are not uncommon in industry. However, as previously discussed in \sect\ref{sect:problem-adversary} the capability to make educated guesses is sufficient, as long as there is a degree of overlap between the guessed labeled samples $\labeledDataset_{\adversary}$ and the actual labeled samples \labeledDataset, i.e., $\labeledDataset_{\adversary} \cap \labeledDataset\not=\emptyset$. In \sect\ref{sect:eval-ablation}, we evaluated scenarios were the adversary's guesses about the labeled images showed only a precision of 20\%, thus, 20\% of the images in $\labeledDataset_{\adversary}$ were actually in \labeledDataset. Further, we performed another experiment where the adversary increased the chance of having at least one correct guess per image by leveraging poisoning patterns that consist of 4 guesses for the labeled images. These experiments demonstrated that the ability to make educated guesses about the labeled dataset with a limited precision is sufficient for \ourname to significantly impact the model’s accuracy.\\
\noindent In addition to the ability to make educated guesses on the labeled dataset, \ourname requires the adversary to be able to upload samples to the attacked platform. It might be a challenge here to control a considerable amount of samples. However, as shown in \sect\ref{sect:eval}, \ourname is effective even with very small Poisoning Data Rates (PDRs). Notably, despite the social networks' significant efforts to identify and block automated content posting, bots continue to successfully upload content automatically~\cite{kirchgaessner23Guardian,woolley2020,metz2020Times,bbc24ukvotes,Tassi2023}, allowing an attacker to post automatically a certain fraction of manipulated media files. Furthermore, Carlini \etal recently demonstrated that significant portions even of established datasets can be manipulated with minimal effort~\cite{carlini2024poisoning}, showing the practical applicability of \ourname.

\noindent Since the attack exploits the training algorithm's ability to predict wrong labels, it is restricted to Semi-Supervised Learning (SSL) settings. It is not applicable for pre-labeled scenarios like self-supervised learning that are used for text processing.

\noindent Further, in line with prior research on adversarial deep learning~\cite{fang2020local,athalye2018synthesizing,liu2018trojaning} and attacks on SSL in particular~\cite{carlini2021poisoning,feng2022unlabeled,franci2022influence}, also for the proposed attack, there is no formal proof to show that the attack works. To the best of our knowledge, all existing literature introducing attacks on SSL algorithms for DNNs show their effectiveness through an empirical evaluation. Consistent with this existing work, in \sect\ref{sect:eval}, we presented an extensive empirical evaluation showcasing the effectiveness of our proposed \ournameAttackNew across diverse datasets, SSL algorithms, input augmentation techniques, and attack parameter variations.

\noindent To craft the poisoning pattern, \ourname selects images with labels different from the original image. Therefore, \ourname requires knowledge on the output space. As observed in App.~\ref{app:eval-output}, the attack is ineffective if the attacker assumes a larger output space and, therefore, more labels than the victim actually uses. This occurs because \ourname uses samples with different labels to create the poisoning pattern, causing the SSL algorithm to mislabel them. However, if \adversary assumes a larger output space than \victim uses, the samples of the poisoning pattern might belong to the same class in \victimGen output space. Notably, the adversary can mitigate this risk by utilizing a reduced output space during the poisoning process. Therefore, the attacker needs to select images from classes that have a very high probability of belonging to different classes. For example, images from machines can be used to poison animal images. As shown in \sect\ref{app:eval-output}, this strategy allows the attack to remain effective even in case of a reduced output space.

\noindent Also, it should noted that the \ournameAttackNew is executed in a blind manner, without any assumptions regarding access or knowledge of the deployed algorithms, training process, or victims. However, this also introduces another challenge, as the adversary \adversary cannot monitor the attacked victim and, thus, cannot verify the success of the attack afterward. Analogous to other attacks against DNNs, such as inference attacks~\cite{shokri2017membership,salem2020updates}, watermark removal strategies~\cite{chen2021refit,wang2019attacks}, or poisoning attacks against self-supervised learning~\cite{shan2024nightshade,shan2023glaze,liang2023adversarial}, the attacker can only measure the attack's effectiveness in a simulated environment. Consequently, it remains infeasible for the attacker to verify the attack's success for the actual victim, therefore, whether it's preventing DNN convergence (\ournameAttackNew), confirming the utilization of extracted data for training (inference attacks), or validating the removal of an unknown watermark (watermark removal approaches).

\subsection{Future Work}
\label{sect:discussion-future}
In our attack, we combined the original image with the poisoned pattern by performing weighted averaging to obtain the manipulated image. This enables the adversary to control the suspiciousness of the image using the PV parameter. Especially images that were crafted using low PV values seem to be inconspicuous to humans, particularly when they are displayed among many other benign images to the users of social media platforms. However, the question of what kind of adversarial perturbations are visible to humans is an active research topic~\cite{schneider2023dual,harding2018human}. In this work, we focused on developing a scheme that combines two images and causes the SSL algorithm to use one for guessing the labels and the other for the actual training, resulting in mislabeled training samples. Therefore, the question on what kind of general patterns in images cause attention by humans or which combination of colors is prone to be easily-detectable is out of the scope of this work. Future work might improve the developed scheme and reduce the number of pixels, e.g., by removing the background of the labeled images that are used for the poisoning pattern. In this context, it should be emphasized that the adversary can still inspect the images before uploading them to ensure that they are inconspicuous. 

\subsection{Potential Counter Measures}\label{sect:discussion-defenses}
In this paper, we introduced the \ournameAttackNew that disrupts SSL training algorithms by uploading manipulated samples to the platform that is used as data source. We demonstrated the risks posed by \ourname through a case-study on three real-world social networks (see \sect\ref{sect:eval-casestudy}). An important research direction for future research is, therefore, to develop effective mitigation schemes against this type of attack.

\noindent We demonstrated in App.~\ref{app:eval} that \ourname is robust against standard image augmentation techniques. Other defense strategies could involve adapting techniques that were developed against adversarial examples. Intuitively, the poisoning pattern that \ourname injects in the manipulated images might be comparable to the noise introduced by adversarial examples. Zantedeschi \etal propose applying Gaussian noise to the images to mitigate adversarial examples~\cite{zantedeschi2017efficient}. However, as we demonstrated in App.~\ref{app:eval}, although applying Gaussian noise reduces the models' performance, \ourname remains effective. Pang \etal\xspace propose a thresholding test based on a kernel density function to detect adversarial examples~\cite{pang2018towards}. However, as shown in \sect\ref{sect:eval-aeDefenses}, the obtained scores for manipulated images are in the same range as for benign images. Also, a dedicated classifier was unable to detect the manipulated images (see \sect\ref{sect:eval-aeDefenses}). A key difference in an adversarial example that might be the reason for the ineffectiveness of these methods might be the fundamental difference in the patterns' structures. Adversarial examples exploit specific DNN parameter values to cause mispredictions, resulting in a random-like poisoning pattern. In comparison, the pattern of \ourname consists of real samples, thus showing structures and edges as regular samples do. Thus, further research is required to investigate potential countermeasures.

\noindent An important aspect for countermeasures against \ourname is the overlay of different samples and that the poisoning pattern is usually only weakly embedded (using low PV values). In the following, we describe two options that future work might investigate to remove the poisoning pattern by exploiting the overlay.\\
\textbf{Noise Reduction Techniques.} Signal processing methods, such as noise reduction, could be utilized for the removal of the poisoning pattern. A challenge here is to erase the pattern without affecting the algorithm's ability to process benign samples. Our evaluation demonstrated that simple compression and Gaussian smoothing are insufficient (see App.~\ref{app:eval}) to mitigate the \ournameAttackNew. Future research might, therefore, focus on more sophisticated signal-processing techniques, such as  singular value decomposition~\cite{sadasivan1996svd}, analyzing the samples in the frequency domain~\cite{hassanpour2008time}, or statistical analysis~\cite{martinez2022algorithms}.\\
\noindent\textbf{Modeling Attack as Cocktail Party Problem.} The challenge of extracting the original sample from overlaid data in our scenario is similar to the well-known cocktail party problem. This problem describes a situation in which multiple audio signals overlap, similar to multiple conversations occurring simultaneously in a crowded room. Different approaches where proposed to separate signals using differential beamformers~\cite{chen2017cracking}, transformer networks~\cite{subakan2021attention}, or asynchronous fully recurrent convolutional neural networks~\cite{hu2021speech}. Leveraging and adapting these techniques might provide a promising research direction to isolate and effectively separate the injected poisoning patterns from the original image while preserving the integrity of the original training samples.

\section{Related Work}
\label{sect:sota}

\noindent In the following, we analyze existing work to attack SSL and untargeted attacks against deep learning. Despite the limited research on adversarial SSL, previous approaches focused on either injecting a backdoor into the resulting model or reducing the accuracy under the assumption that the adversary controls the labeled dataset. However, assuming control of the labeled dataset is not practical as it requires the adversary to control the victim. First, we will discuss existing attacks on SSL (\sect\ref{sect:sota-ssl}), before describing several untargeted attacks against other, non-SSL, learning scenarios~(\sect\ref{sect:sota-other}).

\subsection{Attacks on SSL}
\label{sect:sota-ssl}

\noindent In previous research, various targeted poisoning attacks against SSL have been proposed, such as the white box backdoor attacks developed by Yan \etal~\cite{yan2021deep,yan2021dehib} and Feng \etal~\cite{feng2022unlabeled}, which manipulate samples in the unlabeled dataset to cause the SSL algorithm to make incorrect label predictions. They assume that the adversary can use the parameters of an intermediate version of the DNN model to determine a perturbation that causes the SSL algorithm to guess a wrong label for this specific sample. However, these assumptions are not practical, as this requires the adversary to be able to monitor the victim and the ongoing training process. However, if the adversary has access to the victim and can inspect the intermediate DNN model, there is no need to poison the unlabeled dataset but it can also poison, e.g., the labeled dataset. Neither is it practical to assume that the victim downloads the data a second time during the training. In comparison, \ourname does not require the adversary to have any control or any knowledge about the victim (parameters of the intermediate model, SSL algorithm, hyperparameters). It crafts the manipulated images once, before it is sufficient to upload the manipulated samples to the platform before the training is started~(see \sect\ref{sect:problem-adversary}).

\noindent Other studies~\cite{carlini2021poisoning,connor2022rethinking}, focused on poisoning the unlabeled dataset by adding interpolations between samples from different classes to inject a backdoor into the trained DNN model. However, these attacks restrict the attack objective since they can only cause mispredictions for individual samples, rather than reducing the overall performance of the model (see \sect\ref{sect:eval-ablation}). In contrast, \ourname significantly decreases the utility of the trained model and reduces its performance to that of a naive classifier, making the poisoned dataset unusable for SSL.

\noindent Franci \etal~\cite{franci2022influence} and Liu \etal~\cite{liu2019unified} perform both untargeted poisoning attacks by altering samples from the labeled dataset. They aim to minimize the number of labels that need to be changed to decrease the effectiveness of the models. However, this approach requires the attacker to have access to the labeled dataset, which implies a high level of capability for the attacker. An attacker that is able to control the small and manually labeled dataset of the victim is highly likely to be also able to interfere with the training process and directly sabotage the DNN model. In contrast, our proposed method only requires the attacker to upload manipulated samples to the source of the unlabeled dataset, such as a social network. Therefore, \ourname enables even a weak attacker to carry out poisoning attacks on semi-supervised learning and discourages the use of public data sources for SSL training.

\subsection{Untargeted Poisoning Attacks against Other Learning Scenarios}
\label{sect:sota-other}

\noindent In addition to SSL, untrusted data is also used in online learning scenarios, such as leveraging user feedback for predictions~\cite{pang2021accumulative}. Several approaches have been proposed to address poisoning attacks in these scenarios by treating them as an optimization problem and creating samples that maximize the impact of the attack~\cite{pang2021accumulative,zhang2020online,huang2020metapoison}. However, for online learning systems, the attacker can provide incorrect labels to confuse the system, and also has knowledge about the victim models, such as hyperparameters or even the weights. In contrast, \ourname is a black-box attack, in which the attacker has no knowledge of the DNN model or training algorithm.

\noindent  Federated Learning (FL) is a distributed training approach where different clients perform the training locally and only share the parameters of their DNN models with a coordinating server~\cite{mcmahan2017communication}. The decentralized structure prevents the server from inspecting the training data, thus making it vulnerable to untargeted poisoning attacks by malicious clients~\cite{fang2020local,blanchard2017machine}. However, these attacks typically involve knowledge of the current model's state, such as the current model parameters, and are, therefore, white-box attacks. Also, due to the decentralized training structure, the adversary can manipulate the model's parameters directly rather than being restricted to adding a few samples to the dataset as \ourname does.

\section{Conclusion}

\noindent In this paper, we presented \ourname, a novel untargeted-poisoning attack that disrupts Semi-Supervised Learning (SSL) training by introducing manipulated samples into the unlabeled dataset. While existing attacks on SSL either focus on backdoor attacks or require knowledge about the attacked training process, \ourname operates blindly. It crafts manipulated images before the training process starts without any knowledge about the training algorithm or hyperparameters. \ourname employs a combination of techniques that cause SSL algorithms to overlook the sample's actual content and instead rely on maliciously crafted patterns superimposed on real samples, leading to incorrect label guesses. Through a comprehensive evaluation, we demonstrated \ournameGen effectiveness on several state-of-the-art SSL algorithms, showing that it can be successful with small percentages of manipulated examples. Furthermore, we conducted a real-world case study to illustrate the vulnerability of data obtained from social networks, \mbox{including Facebook, Instagram, and Pinterest, to \ourname.}

\noindent Our work highlights the need for robust training algorithms that can effectively counteract untargeted-poisoning attacks in the context of SSL. Additionally, it emphasizes the importance of considering such attacks in the design and deployment of DNNs, particularly in sensitive domains such as social networks. Our results indicate that untargeted poisoning attacks on SSL can be effective and pose a serious threat to the security of DNNs. Therefore, it is crucial for researchers and practitioners to consider these types of attacks and develop appropriate countermeasures to secure DNNs.%\pagebreak

\section*{Acknowledgments}
\noindent This research received funding from the Horizon program of the European Union under grant agreements No. 101093126 (ACES) and No. 101070537 (CROSSCON), as well as the Federal Ministry of Education and Research of Germany (BMBF) within the IoTGuard project.

\bibliographystyle{ACM-Reference-Format}
\bibliography{main}

\appendix

\section{Image Augmentation}
\label{sect:background-augmentation} 

\noindent Data augmentation is a technique that modifies the individual samples of training data, being used, e.g., to create a larger dataset for training DNNs.

One common example of data augmentation is the use of image mirroring, where the image is flipped horizontally. This technique is particularly useful in increasing the diversity of the training dataset and thereby increasing the robustness and generalization ability of the DNN. By requiring the DNN to learn features that are invariant to horizontal flipping, the model is encouraged to focus on important features of the image and not on the specific orientation of the image. This in turn, improves the model's ability to generalize to new, unseen images. Additionally, this technique also increases the number of training examples and can help reduce overfitting.
As we will further explain in \sect\ref{sect:background-ssl}, certain algorithms for SSL utilize data augmentation to improve the DNN's generalization ability~\cite{devries2017improved, xie2020adversarial, cubuk2019autoaugment}. We categorize data augmentation techniques into two groups: \\

\textbf{Weak augmentation} 

techniques make minimal modifications to the training samples, such that the colors and shapes are preserved. Examples of such techniques include horizontally flipping the images along the center axis and randomly shifting the images by a small number of pixels. \\
\textbf{Strong augmentation} 

involves significant modifications, such as altering the color or obscuring parts of the image. One example of a strong data augmentation technique used by SSL algorithms is RandAugment~\cite{cubuk2020randaugment}, which employs techniques such as image inversion and partial image occlusion. These modifications can make it challenging even for humans to correctly identify the class of the augmented image~\cite{cubuk2019autoaugment}.

\section{Ethical Considerations}
\label{app:eval-ethics}

\noindent In \sect\ref{sect:eval-casestudy}, the effectiveness of our attack on data that were obtained from different real-world social networks is demonstrated. To conduct these experiments, manipulated images were uploaded to three different social networks: Facebook, Instagram, and Pinterest. The effectiveness of our attack was evaluated using the manipulated data from these social networks. To ensure that these experiments do not have any negative impact on anyone, careful consideration and adaptation of the privacy settings for the profiles on each social network were applied. We created on Facebook a private album that only we could access, on Instagram a private account with no followers, and on Pinterest a secret board was created.

\noindent In the present study, particular attention was given to the privacy settings on three popular social media platforms, namely Facebook, Instagram, and Pinterest. For Facebook, the "Select audience" option in the privacy settings was configured to "Only me," thereby ensuring that the visibility of the posted content was limited to the account holder alone. For Instagram, the account policy was set to "Private Account" and the account was not connected to any other account, thereby ensuring that the content was not accessible to others. On Pinterest, the option to "Hide your profile from search engines (e.g., Google)" was enabled, and a secret board was created to upload all images. Additionally, none of the accounts was connected to any other account, thus ensuring that the images remained private.

\noindent As a precautionary measure, the content and corresponding user accounts were immediately deleted after downloading them. Each of the social networks stated that deleting the account would also result in the deletion of any stored data. Furthermore, we carefully analyzed the usage terms of each of the social networks utilized in the study and ensured that conducting the experiments did not violate them. The experiments were also aligned with the ethical guidelines of the involved research institutions.

\begin{figure}[b]
    \centering
    \includegraphics[width=\linewidth]{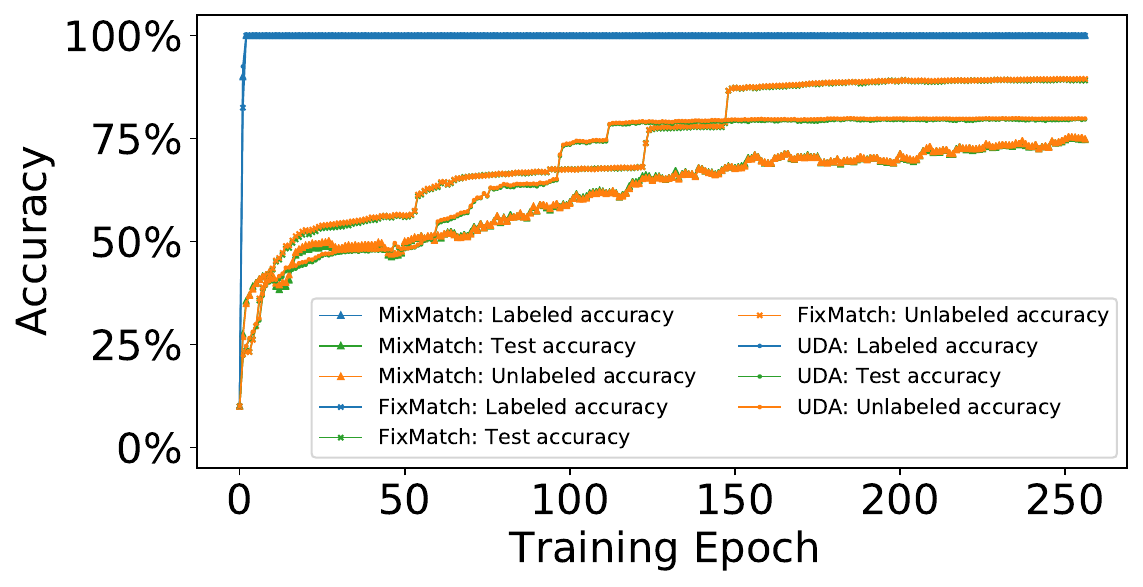}
    \caption{Performance of the DNN for the labeled dataset \labeledDataset, unlabeled dataset \unlabeledDataset and test dataset during training.}
    \label{fig:labeledEvaluation:labeled}\vfill
\end{figure}
\section{Focus of SSL Algorithms on Labeled Dataset}
\label{app:labeledEvaluation}

\noindent The focus of SSL algorithms on the labeled dataset during the early stages of training is an important factor that is exploited in our proposed untargeted-poisoning attack. When training begins with randomly initialized parameters, the guessed labels have low accuracy, making the labeled samples the only source of truth. This results in the DNN overfitting on these samples. Figure~\ref{fig:labeledEvaluation:labeled} shows the accuracy on the labeled dataset, the unlabeled dataset\footnote{For the accuracy calculation we used the labels for those images, which are unknown to the training algorithm.}, and the test dataset during training for all three SSL algorithms (\mixmatch, \uda, \fixmatch) in a scenario without attack. As the figure shows, all three algorithms achieve an accuracy of 100\% on the labeled dataset within 4 epochs, while the accuracy on the unlabeled dataset takes significantly longer to converge.

\noindent Our proposed attack takes advantage of this observation by hiding parts of labeled images in some unlabeled images, which causes the overfitted DNN to mispredict the labels for these images based on the hidden labeled images. By exploiting the focus of SSL algorithms on the labeled dataset, \ourname is able to disturb the training process and lead to incorrect label guesses.

\section{Analysis of Model Behavior}
\label{app:gradcam}

\begin{figure}[tb]
    \centering

     \hfill
     
     \begin{subfigure}[b]{0.135\textwidth}
         \centering
         \includegraphics[width=0.9\textwidth]{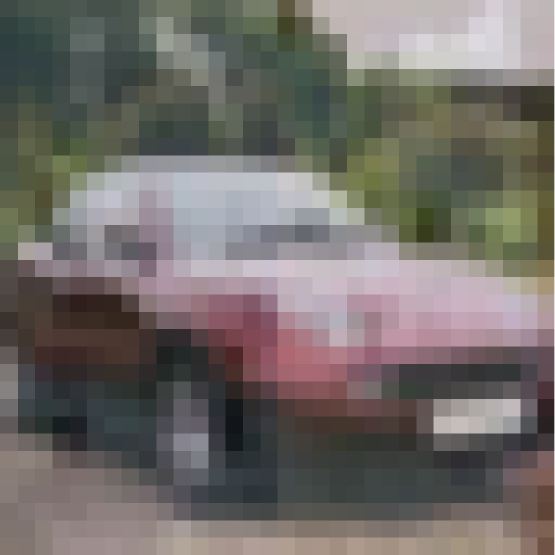}
         \caption{PV = $0.3\%$}
  	    \label{fig:gradcam:manipulatedimage}
     \end{subfigure}
     \hfill
     \begin{subfigure}[b]{0.135\textwidth}
         \centering
         \includegraphics[width=0.9\textwidth]{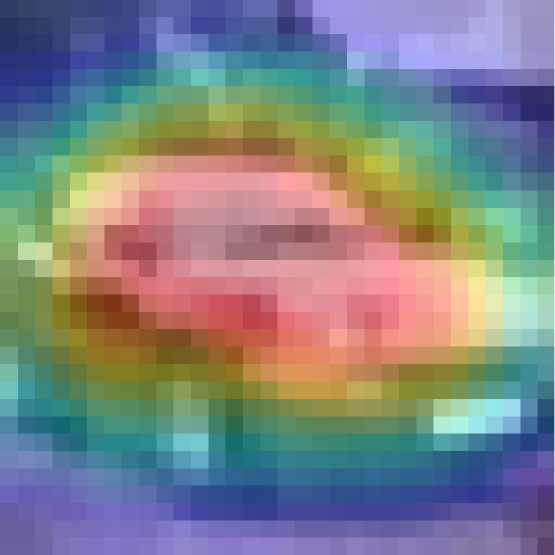}
         \caption{Benign model}
  	    \label{fig:gradcam:normalmodel_manipulatedimage}
     \end{subfigure}
     \hfill
     \begin{subfigure}[b]{0.135\textwidth}
         \centering
         \includegraphics[width=0.9\textwidth]{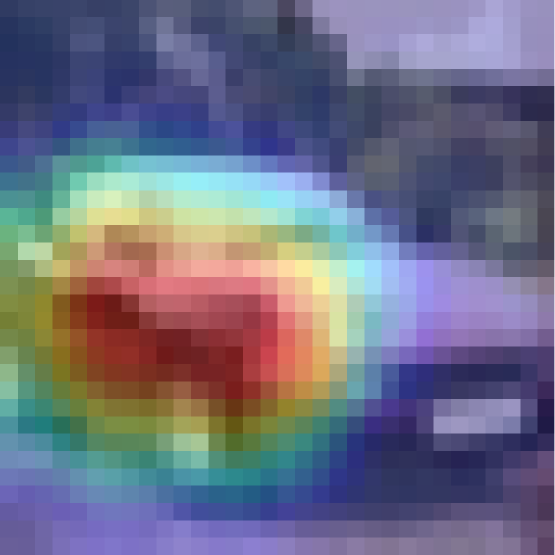}
         \caption{Poisoned model}
  	    \label{fig:gradcam:poisonedmodel_manipulatedimage}
     \end{subfigure}
     \hfill

    \caption{GradCAM of a benign and a poisoned model.}% 
    \label{fig:gradcam}
\end{figure}

\noindent In order to thoroughly analyze how the \ournameAttackNew\xspace manipulatesthe model's behavior, we analyze the saliency map, which illustrates the model's attention. This methodology enables us to compare the model's attention between the benign and poisoned models, thereby providing insight into the attack's impact on the model. For plotting the saliency maps, we employed the widely used GradCam approach~\cite{selvaraju2017grad} and utilized the implementation of Gildenblat \etal~\cite{jacobgilpytorchcam}.

\noindent Figure~\ref{fig:gradcam} shows the saliency maps for a benign and a poisoned model. The saliency map of the benign model reveals that the model's decision-making process is based on the analysis of the whole object, as can be observed in Fig.~\ref{fig:gradcam:normalmodel_manipulatedimage}. In contrast, the saliency map of the poisoned model illustrates the model's attention is primarily focused on the image's right side, where one of the labeled images for the poisoning pattern is located. This can be attributed to the fact that the poisoned model was trained on both the pattern and the car image, while the benign model was only trained on recognizing the car. As a result, the center of attention is shifted to the left for the poisoned image, indicating overfitting on the data. The shift in attention highlights the impact of the poisoning pattern in the manipulated images and \ournameGen effectiveness.

\section{Choice of Density Functions in Backdoor Adaption}
\label{app:density}
In the past, different security attacks on SSL have been proposed~\cite{carlini2021poisoning, connor2022rethinking} that aim to inject a backdoor into the model. In comparison to these attacks, does not inject additional functionality into the model but disturbs the ability to utilize the non-manipulated data, which creates several significant challenges, as discussed in \sect\ref{sect:problem-requ}. In \sect\ref{sect:eval-ablation}, we compared the \ourname attack against different baseline attacks, including an adaption of the attack proposed by Carlini~\cite{carlini2021poisoning}. The attack interpolates between samples of two classes that the attack aims to connect, i.e., making the model classify samples of one class as samples of the second class. In Tab.~\ref{tab:ablationstudy}, we showed only the results for the density function $1.5-x$, that is, according to Carlini~\cite{carlini2021poisoning} the most effective density function. For the sake of completeness, we show in Tab.~\ref{tab:density} the results for other density functions. As the table shows, \ourname is more effective than any of these density-function-based attacks, while the density functions works as regularization and, in consequence, sometimes even remove the model's utility.

\begin{table}[tb]
    \caption{Effectiveness of density functions for an untargeted version of the attack proposed by Carlini~\cite{carlini2021poisoning} for a PDR of 10\%.}
    \label{tab:density}
    \centering
    \scaleTable{
    
        \begin{tabular}{l|rrr}
        Density Function & \uda\\\hline
        Benign Scenario &  79.64\% \\
        Remove 10\% of Samples& 77.01\%\\
        \hline
        %One-Labeled-One-Unlabeled  & 87.02 \\
        $1-x^2 + 0.5$ & 82.47\%\\
        1 & 85.96\%\\
        $1-x$ & 82.92\%\\
        $1.5-x$ & 81.03\%\\\hline
        \ourname (PV=0.2, PDR=10\%) & 67.15\% \\
        \ourname (PV=0.1, PDR=5\%) & 68.71\% \\
        \end{tabular}
    }
\end{table}

\section{Further Evaluation}
\label{app:eval}

\noindent In real-world applications, typically, data preprocessing techniques might be applied to improve computation efficiency, e.g., through data compression to reduce necessary storage size. Other preprocessing techniques can involve image augmentation to improve the performance of the trained model. To evaluate the practical applicability of \ourname, we, therefore, conducted several experiments involving various data-augmentation methods (Gaussian smoothing, JPEG compression, and random image rotation) to asses \ournameGen robustness against these techniques. The evaluation is conducted with a PDR of 5\% and a PV of 0.1. As the results in Tab.~\ref{tab:defenses} show, although the data augmentation impacts model performance even in the absence of an attack, \ourname is still able to reduce the model's utility. Notably, image flipping, an augmentation technique that horizontally flips the original image, is already part of the weak augmentation methods applied in SSL algorithms. Consequently, this technique was included in all our experiments, and no additional tests were conducted for it.

\noindent Furthermore, we assessed the effect of compression on \ourname as part of our case study (see \sect\ref{sect:eval-casestudy}). The reason for this assessment is that compression can affect the quality of the images and, thus, the performance of the algorithms that operate on them. By evaluating the effect of compression on \ourname, we aimed to investigate its robustness to this kind of degradation of the input data.

\begin{table}[b]
    \caption{Effectiveness of \ourname for the \cifar dataset for different data augmentation techniques.}
    \label{tab:defenses}
    \centering
    \scaleTable{
    
        \begin{tabular}{l|rr|rr|rr}
        
        \multirow{2}{*}{} &\multicolumn{2}{c|}{\fixmatch} &\multicolumn{2}{c|}{\mixmatch} &\multicolumn{2}{c}{\uda}\\
        Augmentation &Benign & Attack &Benign & Attack&Benign & Attack\\\hline
        No Augmentation  &  89.10 & 83.68&74.75&64.85& 79.64&68.71\\\hline
        Gauss. Smoothing & 91.98 & 76.21 & 71.45& 68.43&79.53&70.53\\
        Random Rotation & 82.99 & 72.05 & 63.82&61.95&83.28&77.58\\
        JPEG Compression & 87.75 & 74.12&74.25&69.92&83.05&73.48\\
     
        \end{tabular}

    }
\end{table}

\noindent Table~\ref{tab:lowpdrsApp} shows the effectiveness of \ourname for very low PDRs. As the table shows, even for small PDRs, \ourname reduces the accuracy by 10\% for \fixmatch and 5\% for \uda and \mixmatch. Thus \mixmatch and \uda seem to be more robust against very small fractions of manipulated data. Notably, for a PDR of 3\%, the accuracy is reduced by almost 10\%, showing that \ourname is still effective.
\begin{table}[b]
    \caption{Effectiveness of \ourname for very low Poisoned-Data-Rates (PDRs) for the \cifar dataset.}
    \label{tab:lowpdrsApp}
    \centering
    \scaleTable{
    
        \begin{tabular}{l|rrr}
        \multicolumn{1}{c|}{PDR}               & \multicolumn{1}{c}{\fixmatch} & \multicolumn{1}{c}{\uda} & \multicolumn{1}{c}{\mixmatch}\\\hline
        0.0\% (No-Attack) & 89.10 \%  & 79.64 \%  & 74.75 \% \\
        0.1\%             & 77.78 \%  & 73.93 \%  & 72.71 \% \\
        0.5\%             & 79.43 \%  & 72.41 \%  & 69.83 \% \\ 
        3.0\%             & 78.74 \%  & 72.66 \%  & 65.13 \% \\
        5.0\%             & 83.68 \%  & 68.71 \%  & 64.85 \% \\
        \end{tabular}
    }
\end{table}
\vfill
\end{document}